\documentclass[12pt]{article}
\usepackage[mathscr]{eucal}
\usepackage{epsfig,amsfonts}
\usepackage[fleqn]{amsmath}
\usepackage{amsthm,amssymb}
\usepackage{graphicx}
\usepackage{hhline}
\usepackage{cite}
 
\makeatletter
\@addtoreset{equation}{section}
\makeatother
\renewcommand{\theequation}{\thesection.\arabic{equation}}
 
\def\be{\begin{equation}}
\def\ee{\end{equation}}
\def\bdm{\begin{displaymath}}
\def\edm{\end{displaymath}}
\def\bea{\begin{eqnarray}}
\def\eea{\end{eqnarray}}

\def\zb{{\bar z}}
\def\wb{{\bar w}}

\def\ri{{\rm i}}
\def\Xint#1{\mathchoice
    {\XXint\displaystyle\textstyle{#1}}%
    {\XXint\textstyle\scriptstyle{#1}}%
    {\XXint\scriptstyle\scriptscriptstyle{#1}}%
    {\XXint\scriptscriptstyle\scriptscriptstyle{#1}}%
    \!\int}
\def\XXint#1#2#3{{\setbox0=\hbox{$#1{#2#3}{\int}$}
    \vcenter{\hbox{$#2#3$}}\kern-.5\wd0}}

\def\dashint{\Xint-}

\newcommand{\p}{\partial}

\newcommand{\rd}{\mbox{d}}
\newcommand{\re}{\mbox{e}}

\begin{document}

\begin{titlepage}
\begin{flushright}
RU-NHETC-2011-9\\
\end{flushright}

\begin{center}
\begin{LARGE}

{\bf Critical values of the  Yang-Yang functional in the  quantum sine-Gordon model}

\end{LARGE}
\vspace{1.3cm}

\begin{large}

{\bf  Sergei  L. Lukyanov}

\end{large}

\vspace{1.cm}
NHETC, Department of Physics and Astronomy\\
     Rutgers University\\
     Piscataway, NJ 08855-0849, USA\\
\vspace{.2cm}
and\\
\vspace{.2cm}
L.D. Landau Institute for Theoretical Physics\\
  Chernogolovka, 142432, Russia\\
\vspace{1.0cm}

\end{center}

\vspace{0.5cm}

\begin{center}
\centerline{\bf Abstract} \vspace{.8cm}
\parbox{15.5cm}
{
The critical values  of the  Yang-Yang functional corresponding 
to the vacuum states of the   sine-Gordon QFT in the finite-volume are studied.
Two major applications are discussed: (i) generalization
of  Fendley-Saleur-Zamolodchikov
relations to  arbitrary values of the  sine-Gordon  coupling constant,
and (ii)   connection problem for  a  certain two-parameter family of solutions of
the   Painlev${\acute {\rm e}}$ III equation.
}

\end{center}

\vfill

\end{titlepage}
\newpage


\tableofcontents

\section{Introduction}


Throughout the past,
a  number of important facts about the  quantum sine-Gordon model  were discovered.
Among them are   elegant relations between the zero-point energy
 and the    Painlev${\acute {\rm e}}$ III transcendent.
To describe them  explicitly  let us recall  some elementary facts about  a structure of the Hilbert space
of the model, 
\begin{eqnarray}\label{sg}
\mathcal{L}={ \frac{1}{\beta^2_{\rm sg}}}\ \Big(\, \ { \frac{1}{2}}\  (\partial_\mu \phi)^2+\Lambda\ 
\cos(\phi)\, \Big)\ ,
\end{eqnarray}
in  finite-size    geometry with the
spatial coordinate $x$  compactified on
a circle of a circumference $R$, with the  periodic boundary conditions
\bea\label{nasbash}
\phi(x+R,t)=\phi(x,t)\ .
\eea
Due to the periodicity of the potential term $\Lambda\, \cos(\phi)$ in \eqref{sg} in
$\phi$, the
space of states $\mathcal{H}$ splits into orthogonal subspaces
$\mathcal{H}_k$, characterized by the ``quasi-momentum'' $k$,
\begin{eqnarray}\label{quasi}
\phi \to \phi +2\pi \,: \qquad \mid \Psi_k\,\rangle \
\to\ \re^{2\pi \ri\,k}\,\mid \Psi_k \,\rangle
\end{eqnarray}
for $\mid \Psi_k\,\rangle \in \mathcal{H}_k$. We call $k$-vacuum
the ground state of the finite-size system \eqref{sg} in the
sector $\mathcal{H}_k$ and denote it  by   $|\,\Psi^{{\rm (vac)}}_k\,\rangle$.  The corresponding energy will be  denoted by  $E_k$.

In general, the  coupling constant  in \eqref{sg} 
should be restricted by the condition $\beta^2_{\rm sg}<8\pi$ \cite{Coleman:1974bu} and
it is  convenient to substitute $\beta^2_{\rm sg}$ for the ``renormalized  coupling'':
\bea\label{ososao}
\xi=\frac{\beta^2_{\rm sg}}{8\pi -\beta^2_{\rm sg}}\ .
\eea
The value
$\xi=2$ 
is   special. For this coupling,   the theory  possesses  ${\cal N}=2$  supersymmetry
which is spontaneously broken, except  the  subspaces
$\mathcal{H}_k$  corresponding to
$k=\pm\frac{1}{4}$\ \cite{Saleur:1991hk}.
In the  sectors  with unbroken supersymmetry the ground state  energy is  of course identically zero. 
In Ref.\cite{Fendley:1992jy} Fendley and Saleur (see also related Ref.\cite{Cecotti:1992qh})  applied the general 
construction \cite{Cecotti:1991me} 
to  derive the remarkable relation
\bea\label{sssiasai}
\frac{ R}{\pi}\ \Big(\frac{\partial E_{k}}{\partial k} \Big)_{\xi=2\atop k=\pm 1/4}=\mp 4 r\,
 \frac{\rd U(r)}{\rd r}\ .
\eea
Here  the
variable $r$ stands  for  the size of the system measured in the units of the correlation length
(inverse soliton  mass $M$),
\bea\label{sssat}
r=MR\ ,
\eea
and
$U=U(t)$ is a particular solution    to the   Painlev${\acute {\rm e}}$ III equation
\bea\label{soisiasai}
\frac{1}{t}\ \frac{\rd}{\rd t}\Big(t\,\frac{\rd U}{\rd t}\Big)=\frac{1}{2}\ \sinh(2 U)\ .
\eea
This equation admits a one-parameter family of  solutions  regular at $t>0$  (see e.g.\cite{McCoy:1976cd}) called 
the     Painlev${\acute {\rm e}}$ III  transcendents.
The special solution in \eqref{sssiasai} is fixed by the following boundary conditions
\bea\label{sosias}
U(t)=
\begin{cases}
-\frac{1}{3}\ \log(t)+O(1)\ \ \ \  \  {\rm as}\ \ \ \ t\to 0\\
\ o(1)  \ \ \ \ \  \  \ \ \ \ \  \ \ \ \ \ \ \ \ \ \ \   {\rm as}\ \ \ \ t\to \infty
\end{cases}\ .
\eea
In the  consequent work\ \cite{Zamolodchikov:1994uw}   Alyosha  Zamolodchikov derived one   more mysterious  relation
\bea\label{siasai}
\frac{ R}{ \pi}\ \Big(\frac{\partial E_{k}}{\partial \xi} \Big)_{\xi=2\atop k=\pm 1/4}
=-\frac{r^2}{8}+
\frac{1}{2}\  \int_r^\infty\rd t\ t\ \sinh^2U(t)\ .
\eea
Below, we will refer to  Eqs.\,\eqref{sssiasai},\,\eqref{siasai} as   the FSZ relations.

\begin{figure}
\centering
\includegraphics[width=10  cm]{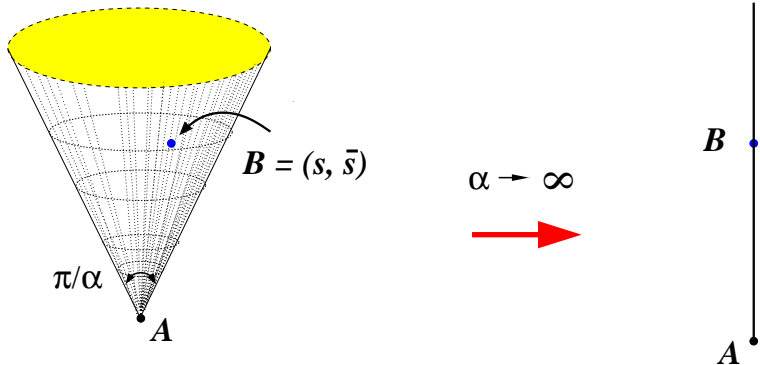}
\caption{ The world sheet for the MShG equation\ \eqref{shgz}.
The dots $A$ and $B$ indicate positions of the apex  and  zero of $p(z)$, respectively. 
At the minisuperspace limit
($\alpha\to\infty$, $r$ is kept fixed)  the world sheet shrinks to a single ray.}
\label{fig0}
\end{figure}

Relations similar  to\ \eqref{sssiasai} and \eqref{siasai}
were also  discovered   in other   models\ \cite{Cecotti:1992qh}, \cite{Fendley:1999zd}, \cite{Bazhanov:2004hv}.
However  all the  generalizations  had  limitations in  the  choice  of  coupling constants and
sectors of  the theories.
The long-time consensus was the
FSZ relations  are 
due  to the  accidental   symmetry
and do not possess any interesting generalizations  for  general values of  $\xi$ and $k$.
The first serious sign that this may not  be  true    came from the  study of $D=4$ supersymmetric  
gauge theories \cite{Gaiotto:2008cd, Gaiotto:2009hg, Alday:2009yn, Alday:2009dv, Alday:2010vh}.
In these  works  a link  was observed     between certain  
Thermodynamic Bethe Ansatz (TBA)  type  integral equations and partial differential equations
integrated  by  the inverse scattering methods.
Some of the  integral  equations  were in fact   identical to
the sine-Gordon TBA systems 
corresponding  to   $\xi\not =2$ and $k\not=\pm\frac{1}{4}$.
Inspired by  this remarkable  development, A. Zamolodchikov and the author found
a classical integrable equation  associated with the quantum sine-Gordon model
for generic $\xi$ and $k$\ \cite{Lukyanov:2010rn}.
It turned out  to be the classical Modified Sinh-Gordon equation (MShG)
\begin{eqnarray}\label{shgz}
\p_z\p_{\bar z}\eta -\re^{2\eta}+p(z)\,{p}({\bar z})\ \re^{-2\eta}=0
\end{eqnarray}
with  $p(z)$ of the form 
\begin{eqnarray}\label{kskssls}
p(z) = z^{2\alpha}-s^{2\alpha}\,.
\nonumber
\end{eqnarray}
Parameters $\alpha$ and $s$ are real and  positive,
related to the
sine-Gordon parameters $\xi$\ \eqref{ososao} and  $r=MR$\ \eqref{sssat}  as follows
\begin{eqnarray}
\label{aoisasos}
\alpha={\xi}^{-1}\ , \ \ \ \ \ \ \qquad\qquad 
s=\Big( \frac{2\,r}{\xi\, r_\xi}\Big)^{\frac{\xi}{1+\xi}}\ ,
\eea
where, for future references, we use  the notation
\bea\label{tsrars}
r_\xi=\frac{2\sqrt{\pi}\, \Gamma(\frac{\xi}{2})}{\Gamma(\frac{3}{2}+\frac{\xi}{2}) }\ .
\end{eqnarray}
The MShG equation in general has no rotational symmetry.
Instead, it has the discrete
symmetry
$z \to \text{e}^\frac{\ri\pi}{\alpha}\, z\,,  \ 
\zb \to \text{e}^{-\frac{\ri\pi}{\alpha}}\, {\bar z}$.
Solutions of the MShG equation \eqref{shgz} relevant to the problem
respect   this symmetry, are  continuous at all finite nonzero $z$, and
grow slower then the  exponential as $|z|\to\infty$.
In other words, they  are single-valued functions
on a cone  with the apex angle $\frac{\pi}{\alpha}$ including the zero of $p(z)$ (see Fig.\ref{fig0}).
There is a one-parameter family  of such solutions, characterized
by the behavior at the apex:
$\eta \to 2l\,\log|z|+O(1)$ as $|z|\to 0$,
with
real $l\in (-\frac{1}{2},\,\frac{1}{2}\, )$ which  turns out to be related to the
quasi-momentum\ \eqref{quasi} by
\bea\label{llsslasa}
l=2\,|k|-{\textstyle\frac{1}{2}}\ .
\eea

The MShG equation  
can be represented  as a flatness condition
for certain  $SL(2)$ connection. In Ref.\cite{Lukyanov:2010rn} it was shown that the
monodromy from  the apex  to infinity  corresponding to the above described solution
is essentially the  sine-Gordon  $Q$-function,
whose asymptotic expansions 
generate the vacuum eigenvalues of integrals of motions
of the quantum theory.

The  original motivation for the present  work was to  incorporate  the FSZ relations
to the construction   of Ref.\cite{Lukyanov:2010rn}.
This problem is solved in  Section\,\ref{Sect2} of this work.
It turned  out that the main player in the  derivation of  the generalized FSZ relations
is a properly defined  ``on-shell'' action for the  MShG
equation.
Remarkably   it can   also be interpreted as    a critical value of the 
Yang-Yang (YY) functional in the quantum sine-Gordon model.

In the seminal  work  \cite{Yang}   the variational principle was applied  to prove an
existence of a  solution to  vacuum Bethe Ansatz (BA) equations for the XXZ spin chain.
From that time the functional whose  extremum condition reproduce  BA equations bears the  Yang-Yang name.
The YY-functional  proves to be   useful for  computing      norms of the  Bethe states 
(see e.g. \cite{Korepin} and references therein).
Recently it attracts a great deal of attention in the context  of the  relation between
supersymmetric gauge theories
and  quantum mechanical  integrable systems\ \cite{Moore:1997dj, Gerasimov:2006zt, Nekrasov:2009rc}.
However, the r${\hat {\rm o}}$le of the  YY-functional in  2D  QFT  seems to be undervalued.
To the best of my knowledge, 
it was never defined in
intrinsic  terms of   integrable QFT.
Nevertheless, there is   a  brute-force approach for the  calculation
of  critical  values of the  YY-functional    in   the sine-Gordon   QFT.
It is  based on the
discretization of the theory, i.e., reducing it to the system with finite number of degrees of freedom,
which then
can be solved by  standard methods of BA.
Although this  formal  approach  does  not clarify  the meaning
of the YY-functional itself, it is sufficient for the calculation of 
the YY-functions, i.e., the  critical values of  YY-functional corresponding to 
the Bethe states.
In this work we restrict  our attention to the $k$-vacuum state   $\mid \Psi^{{\rm (vac)}}_k\,\rangle \in \mathcal{H}_k$.
In Section\,\ref{Sect3} it is shown that the corresponding YY-function
can be identified with the  on-shell   action for the  MShG equation.
Another purpose of  Section\,\ref{Sect3} is to discuss   technical tools for the  calculation of  the YY-function.
We   review   the well-known approach\ \cite{ Klumper:1991, Destri:1992qk} which allows one to
express  partial  derivatives of  the  YY-function   in terms of  a  solution
to  the non-linear  integral  equation from   Ref.\cite{Destri:1992qk}.

Section\,\ref{SectionMini}  is devoted to  the  so-called minisuperspace approximation
(in  the stringy terminology).
The approximation implies the   $\xi\to 0$ limit  such that the soliton mass $M$ is kept  fixed.
In this case
the sine-Gordon QFT reduces to the quantum mechanical problem of
particle in the cosine 
potential.\footnote{Note that in   the conventional
classical  limit,   the mass  of the lightest particle in  \eqref{sg}  is kept    fixed  while $M\to\infty$.}
At the minisuperspace limit the world sheet of  the MShG equation  collapses into a single ray (see Fig.\,\ref{fig0}) 
and the  solution $\eta$  at the segment $(A,\,B)$ is expressed in terms of
a solution of  the Painlev${\acute {\rm e}}$ III equation \eqref{soisiasai},
subject of the  boundary conditions
\bea\label{ytrsosias}
U(t)=
\begin{cases}
\ 2l\, \log(t)+O(1)   \, \ \ \ \  \ \ \  \  {\rm as}\ \ \ \ t\to 0\\
-\log(r-t)+O(1)\ \ \ \  \  {\rm as}\ \ \ \ t\to r
\end{cases}\ .
\eea
Real  solutions of the Painlev${\acute {\rm e}}$ III equation which are  regular at the open segment  $t\in (0,r)$,
and  satisfy 
the boundary conditions \eqref{ytrsosias}, form a 
family which is  parameterized by $r>0$ and $-\frac{1}{2}< l<\frac{1}{2}$.
By taking the minisuperspace limit of the generalized FSZ relation, we  solve the connection problem
for the local expansions of the solution  $U(t)$\ \eqref{ytrsosias} at $t=0$ and $t=r$. 
The results obtained in Section\,\ref{SectionMini}   provide  
 an interesting link between    the    Painlev${\acute {\rm e}}$ III and Mathieu   equations.

\section{\label{Sect2}On-shell action for the ShG equation}

\subsection{From MShG to ShG}

\begin{figure}
\centering
\includegraphics[width=12  cm]{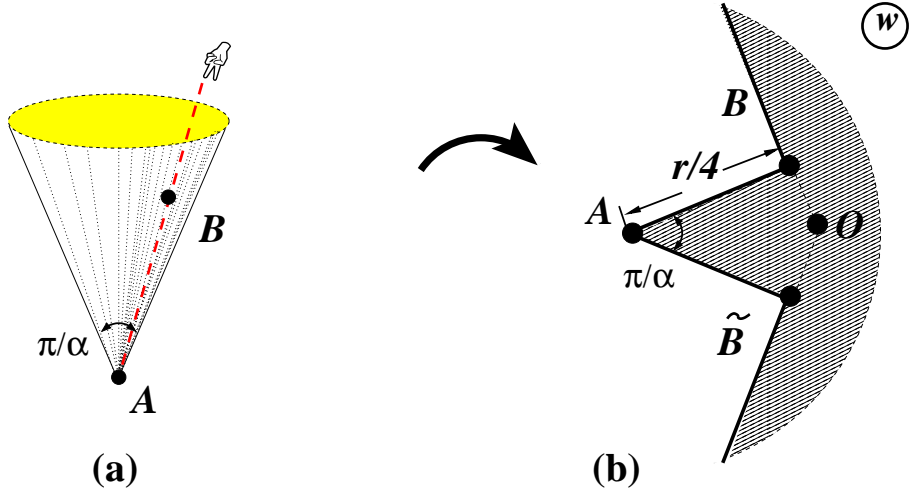}
\caption{
The $w$-image of the cutted  cone  
under the map \eqref{uyslaskasa}.
The points on the cone and their images are denoted by the same symbols.
The segment $A{\tilde B}$ is identified
with $AB$, and the boundary line from ${\tilde B}$ to infinity is identified with the line
from $B$ to infinity. The point $O$ is an origin of the $w$-plane.}
\label{fig1a}
\end{figure}

In practical calculations
it is convenient
to trade the world sheet variable $z$ in the MShG equation
\eqref{shgz} to
\bea\label{uyslaskasa}
 w=\re^{\frac{\ri\pi(\alpha+1)}{2\alpha}}\ \int \rd z\  \sqrt{p(z)}\ ,
\eea
and similarly for $\wb$. 
The  branch of the multivalued function\ \eqref{uyslaskasa}
can be chosen   to  provide the map of the cone with the cut  along the ray $(AB)$ visualized in Fig.\ref{fig1a}a to the
domain of the $w$-complex plane  in Fig.\ref{fig1a}b\ (see Ref.\,\cite{Lukyanov:2010rn} for details).
This conformal map  brings the MShG equation
to the conventional Sinh-Gordon (ShG) form 
\bea\label{luuausay}
\partial_w{ \partial}_{\bar w}{\hat\eta}-\re^{2{\hat\eta}}+\re^{-2{\hat\eta}}=0
\eea
for
$ {\hat\eta}= \eta-{\textstyle \frac{1}{4}}\ \log \big(\,p(z)  p({\bar z})\,\big)$, which  vanishes at infinity
\bea\label{fssisaisa}
\lim_{|w|\to\infty}{\hat \eta}=0\ ,
\eea
becoming  singular at  the apex 
\bea\label{skssai}
{\hat\eta}= 2l\ \log|w-w_{A}|+ O(1)\ \ \ \  \  {\rm as}\ \ \ \ \ |w- w_{A}|\to 0
\eea
and at the point  $B\sim {\tilde B}$
\bea\label{skssaisusy}
{\hat\eta}= -\frac{1}{3}\ \log|w-w_{B}|+O(1)\ \ \ \  \  {\rm as}\ \ \ \ \
|w- w_{B}|\to 0\ .
\eea
Unlike  the apex  singularity, the asymptotic  \eqref{skssaisusy} is an artifact of the
coordinate transformation\ \eqref{uyslaskasa}.

\subsection{\label{Section2b}Action functional}

\begin{figure}
[!ht]
\centering
\includegraphics[width=5  cm]{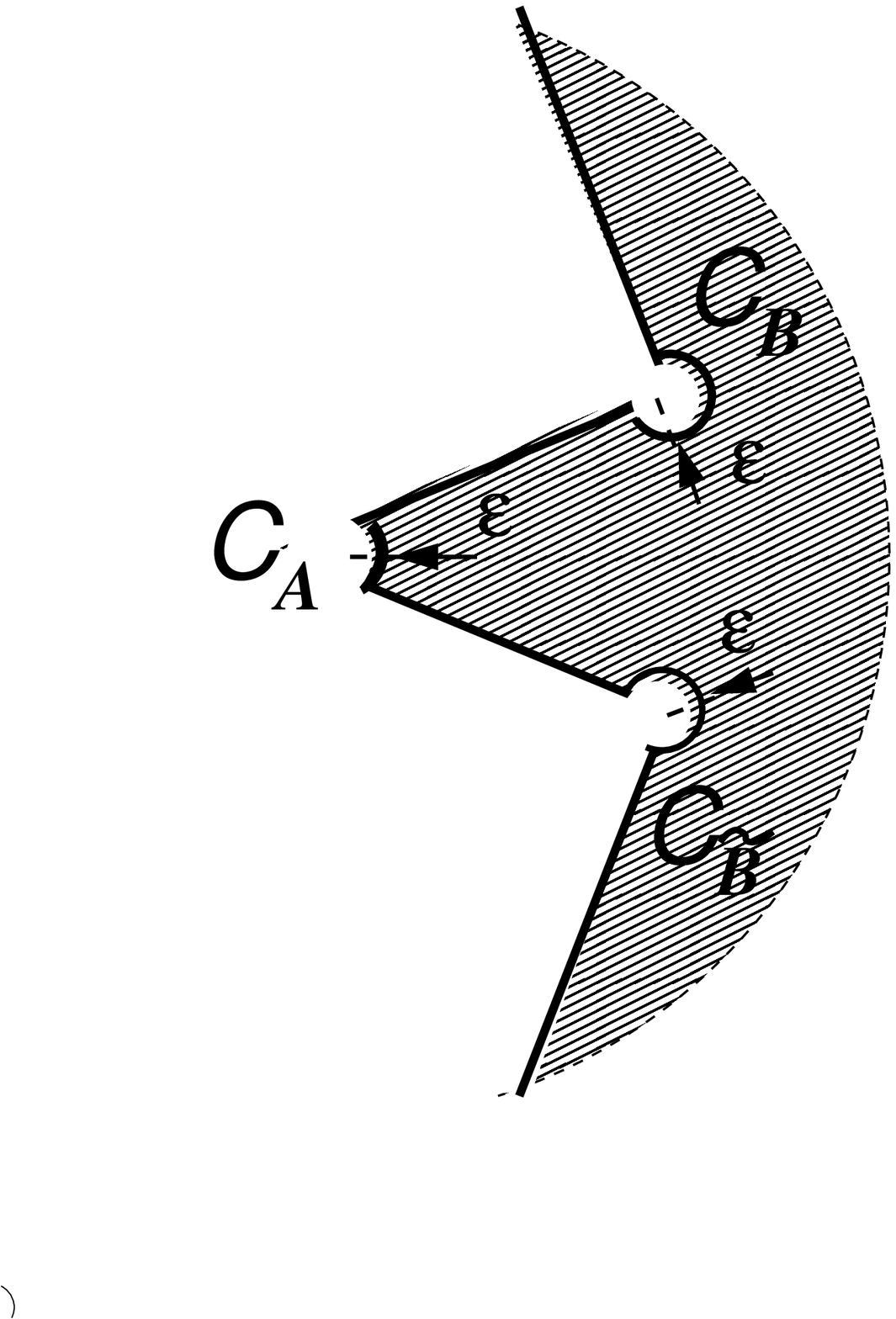}
\caption{The integration domain $D_{\epsilon}$ for the regularized action
\eqref{ssiisa}.}
\label{fig2d}
\end{figure}

To  generalize   relations \eqref{sssiasai},\,\eqref{siasai} we need   
an  extra  ingredient  --  the ``on-shell'' action for the ShG 
equation\ \eqref{luuausay}.
It can be defined  through the following   limiting procedure.
Start with the domain $D$ depicted in Fig.\ref{fig1a}b of the complex $w$-plane. Cut out the small sectors of radius $\epsilon$
around the point $A$, $B$ and ${\tilde B}$ to obtain  the domain $D_\epsilon$  shown  in  Fig.\ref{fig2d}.
Define the regularized  action functional 
\bea\label{ssiisa}
{\cal A}[\,{\hat \eta}\,]&=&\lim_{\epsilon\to 0}\, \bigg[\,  \int_{D_\epsilon} \frac{\rd w \wedge\rd {\bar w}}{2\pi\ri }\
\big(\, \partial_w{\hat  \eta} \partial_{\bar w}{\hat \eta}+4\,\sinh^2({\hat\eta})\,\big)+
\frac{ l}{\pi \epsilon}\ \int_{C_{A}}\rd \ell\
{\hat \eta}-\frac{l^2}{\alpha}\ \log(\epsilon)\nonumber\\
&-& \frac{ 1}{6\pi \epsilon}\ \int_{C_{B}}\rd \ell \
{\hat \eta}-
\frac{1 }{ 6\pi\epsilon}\ \int_{C_{{\tilde B}}}\rd \ell \
{\hat \eta}
-\frac{1}{12}\ \log(\epsilon)\,  \bigg]\ .
\eea
The first term is the ``cutoff'' version of the naive action for  the ShG equation\ \eqref{luuausay}.
The additional terms involves   integrals  over 
three arcs   $C_{A}$, $C_{B}$ and $ C_{{\tilde B}}$
and field-independent counterterms which provide  an existence of the limit.
Then  the ShG equation  supplemented by
asymptotic behaviors near the singularities \eqref{skssai}, \eqref{skssaisusy}
and at large $w$\ \eqref{fssisaisa} constitute a sufficient  condition  for an extremum   of
the   functional \eqref{ssiisa}:
\bea\label{sossai}
\delta {\cal A}=0\ .
\eea
Finally we define  the on-shell action ${\cal A}^*$ as
the value of ${\cal A}[{\hat \eta}]$
calculated on the solution ${\hat \eta}$ \eqref{luuausay}-\eqref{skssaisusy}.

For the variation \eqref{sossai} 
the world sheet geometry, as well as  the parameter $l$  controlling  the behavior  of the solution at the apex,
is  assumed to be  fixed.
Varying     the on-shell  action
with respect to the parameter  $l$, it is observed that
\bea\label{uyososai}
\Big(\frac{\partial {\cal A}^*}{\partial l}\Big)_{r,\alpha}=\frac{1}{\alpha}\ {\hat \eta}_A\ ,
\eea
where the constant ${\hat\eta}_A$ can be thought of  as a regularized value of
the solution  $ {\hat \eta}$ at the apex
\bea\label{eaisusy}
{\hat\eta}_A=\lim_{|w-w_A|\to 0}\big(\, {\hat \eta}(w,{\bar w})-2l\, \log|w-w_A|\,\big)\ .
\eea
It should be stressed that unlike  $l$,  which is the ``input'' parameter applied
with the problem, the value of the constant  ${\hat \eta}_A$ is not prescribed in advance
but determined through the solution, i.e. it is rather part
of the ``output''.

\begin{figure}
\centering
\includegraphics[width=10  cm]{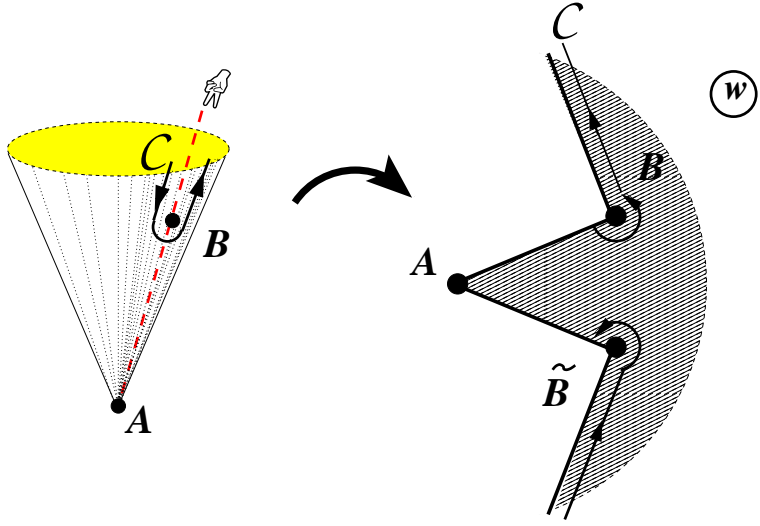}
\caption{ The integration contour $C$
in \eqref{sisisias}.   The contour  on the cone and its  $w$-image
are denoted by the same symbol.}
\label{fig4a}
\end{figure}

Let us consider now   the  infinitesimal variations  of the  world-sheet geometry.
The corresponding $\delta{\cal A}$  do not vanish on-shell
and can  be expressed through   the  on-shell values of the  stress-energy tensor. 
Under 
the   infinitesimal dilation
$\frac{\delta r}{r}=\frac{\delta\epsilon}{\epsilon}=\lambda \ll 1$, 
\bea\label{xsososa}
\delta_r  {\cal A}^*=\frac{\delta r}{r}\, \bigg[\,   
  \lim_{\epsilon\to 0}\int_{D_\epsilon} \frac{\rd w\wedge \rd {\bar w}}{\pi\ri } \  \Theta-\Big(
\frac{ l^2}{\alpha}+\frac{1}{12}\, \Big)\, \bigg]\ ,
\eea
where
\bea\label{hssasys}
\Theta=4 \,\sinh^2({\hat \eta})
\eea
is  a trace of 
the stress-energy tensor for the classical ShG equation.
The other two non-vanishing components of $T_{\mu\nu}$
are given by
\bea\label{sosisia}
T= (\partial_w{\hat \eta})^2\, ,\ \  \ \ \ \ \ {\bar T}= (\partial_{\bar w}{\hat \eta})^2\ .
\eea
By virtue of the  ShG equation, they  satisfy
the continuity equations 
\bea\label{sa]assa}
\partial_{\bar w}T=\partial_w\Theta\ ,\ \  \ \ \ \partial_{ w}{\bar T}=\partial_{\bar w}\Theta\ ,
\eea
and, hence,  they can be  expressed in terms of a single    scalar potential
\bea\label{ystsossai}
T=\partial^2_w\Phi\ ,\ \ \  \ \ {\bar T}={\bar \partial}^2_w\Phi\ ,\ \ \ \ \  \
\Theta=\partial_w\partial_{\bar w}\Phi\ .
\eea
Combining the last formula with\ \eqref{xsososa}, one has
\bea\label{isxssosa}
r\, \Big(\frac{\partial  {\cal A}^*}{\partial r}\Big)_{\alpha,l}=
 \lim_{\epsilon\to 0}\int_{D_\epsilon} \frac{\rd w\wedge \rd{\bar w} }{\pi\ri } \  \partial_w\partial_{\bar w}\Phi
-\Big(
\frac{ l^2}{\alpha}+\frac{1}{12}\, \Big)\ .
\eea
The 2-fold integral here  can be   reduced to the linear integral over the boundary  of  $D_\epsilon$.
The linear  integrals over  the arcs  $C_A$, $C_B$ and $C_{\tilde B}$ 
from  Fig.\ref{fig2d}   cancel out the   term in the brackets in \eqref{isxssosa}.
This follows
from  the asymptotic formulas
\bea\label{oosaosa}
{ \Phi}(w,\,{\bar w})=  -2l^2\  \log|w-w_{A}|+O(1)\ , \ \ \  \  \ \ \ \ \  |w- w_{ A}|\to 0
\eea
and
\bea\label{kssysai}
{ \Phi}(w,\,{\bar w})
=-\frac{1}{18}\times
\begin{cases}
\log|w-w_{B}|+O(1)\ ,\ \ \  \  \ \ \ \ \  |w- w_{ B}|\to 0\\
\log|w-w_{\tilde B}|+O(1)\ ,\ \ \  \  \ \ \ \ \  |w- w_{\tilde B}|\to 0
\end {cases}\  ,
\eea
which are consequences of  Eqs.\,\eqref{sosisia},\,\eqref{ystsossai} and\ \eqref{skssai},\,\eqref{skssaisusy}.
To proceed further, we  need  
to use some properties  of the potential $\Phi$  discussed   in  Appendix\,\ref{AppendixA}.
Namely, for $|w|>|w_B|$
\bea\label{ssisaias}
\Phi\big( w\,\re^{ \frac{\ri \pi (\alpha+1)}{\alpha}},\,  {\bar w}\,\re^{-\frac {\ri \pi (\alpha+1)}{\alpha}}\,\big)=
\Phi(w,\,{\bar w})
\eea
and 
\bea\label{sosssiasai}
\lim_{|w|\to \infty}{ \Phi}(w,\,{\bar w})= 0\ .
\eea
Eq.\eqref{ssisaias}
implies that
the half-infinite boundary  rays $(B,\,\infty)$  and $({\tilde B},\,\infty)$ from  Fig.\ref{fig2d} 
do not  contribute into the integral  \eqref{isxssosa}.
Now, taking into account  Eq.\eqref{sosssiasai}, it is straightforward to show   that
\bea\label{xososa}
r\, \Big(\frac{\partial  {\cal A}^*}{\partial r}\Big)_{\alpha,l}=-\frac{1}{2\pi}\  
\sin\Big({ \frac{\pi}{2\alpha}}\Big)\  \big(\,  {\mathfrak J}_1+ {\bar   {\mathfrak J}}_1\,\big)
\ ,
\eea
where  notations from  Ref.\cite{Lukyanov:2010rn} are adopted,
\bea\label{sisisias}
 \sin\big({\textstyle \frac{\pi}{2\alpha}}\big)\  {\mathfrak J}_1&=&
{\textstyle \frac{1}{4}}\ \   \re^{\frac{\ri (\alpha+1) \pi}{2\alpha}}\ \int_{C}\big(\, \rd  w\, T+\rd {\bar  w}\, \Theta\,\big)\\
 \sin\big({\textstyle \frac{\pi}{2\alpha}}\big)
\  {\bar   {\mathfrak J}}_1&=&
{\textstyle \frac{1}{4}}\ 
\    \re^{-\frac{\ri (\alpha+1) \pi}{2\alpha}}\int_{  C} \big(\, \rd  {\bar  w}\, {\bar T}+ { \rd w}\, \Theta\,\big)\ .\nonumber
\eea
The integration contour $C$ is visualized in
Fig.\,\ref{fig4a}.
Due to the continuity equations,  $ {\mathfrak J}_1$ and ${\bar  {\mathfrak J}}_1$ are  
integrals of motion, i.e.,  they do
not change under continuous deformations of the  integration  contour.

Finally, let us  consider   the variation of  the on-shell ShG action
under an infinitesimal  change  of the apex angle $\frac{\pi}{\alpha}$.
In this case, using the simple electrostatic analogy,  one can express  $\delta_\alpha {\cal A}^*$
through  the torque applied to the boundary $\partial D_\epsilon$  
\bea\label{tyssosiosa}
\delta_\alpha  {\cal A}^*&=&\delta\Big(\frac{\pi}{\alpha}\Big)\ \lim_{\epsilon\to 0} \bigg[\, 
-\int_{\partial D_\epsilon}\frac{\rd \ell}{\pi}\   \epsilon^{\mu\nu} x^\mu n^{\sigma}\ T_{\nu\sigma}+
\frac{l}{\pi}\ {\hat \eta}_A+ \frac{l^2}{\pi}\ \log(\epsilon)\, \bigg]\ ,
\eea
where  $x^{1}=\Re e(w-w_A)$ and $x^2=\Im m (w-w_A)$ are real coordinates
on $D_\epsilon$, $n^\sigma$ is a unit external normal to the boundary $\partial D_\epsilon$
and
$ T_{\mu\nu}= -\frac{1}{4}\  \partial_\mu{\hat \eta}\partial_\nu{\hat \eta}+
\delta_{\mu\nu}\ \big[\, \frac{1}{8}\
\big(\partial_\sigma{\hat\eta})^2+2\, \sinh^2({\hat \eta})\, \big]$.
The integration  contour  $\partial D_\epsilon$ contains
two components $\partial D^{+}_\epsilon$ and $\partial D^{-}_\epsilon$,
related by   reflection on   the  axis  $x^{2}=0$.
Since  each  component contributes equally, we  replace   the  integral  in \eqref{tyssosiosa},
by  $2\int_{\partial D^{+}_\epsilon}$ and  then evaluate it
using  the  identity
\bea\label{saosai}
4\ \int_{C}\rd \ell\   \epsilon^{\mu\nu} x^\mu n^{\sigma} T_{\nu\sigma}
=\int_C \rd x^\mu\, \partial_\mu   \Phi-\int_C\rd \ell\ \partial_\mu \big( x^\mu \Phi)
\eea
and Eq.\eqref{sosssiasai}.
This yields 
\bea\label{issusossa}
\alpha^2\, \Big(\frac{\partial  {\cal A}^*}{\partial \alpha}\Big)_{r,l}
=-{\textstyle \frac{1}{2}}\ \Phi_A
-l\ {\hat \eta}_A\ ,
\eea
where ${\hat\eta}_A$ is given by Eq.\eqref{eaisusy} and   $\Phi_A$ stands for
another ``output'' constant determined through  the solution of the ShG equation -- the  regularized value of
the potential at the apex
\bea\label{isusy}
\Phi_A=\lim_{|w-w_A|\to 0}\big(\, \Phi(w,{\bar w})+2l^2\, \log|w-w_A|\,\big)\ .
\eea

\subsection{Generalized FSZ relations}

The compatibility of  the   derived above   equations \eqref{uyososai},\,\eqref{xososa} and \eqref{issusossa}
implies
\bea\label{sisissa}
\alpha\, \Big(\frac{\partial {\mathfrak F}}{\partial l}\Big)_{r,\alpha}&=&- r\,\Big(\frac{\partial{\hat \eta}_A}{\partial r}\Big)_{\alpha,l}\\
\alpha^2\, \Big(\frac{\partial {\mathfrak F}}{\partial \alpha}\Big)_{r,l}&=&
{\textstyle \frac{1}{2}}\ 
r\,\Big(\frac{\partial{\Phi}_A}{\partial r}\Big)_{\alpha,l}
+l\, r\,\Big(\frac{\partial{\hat \eta}_A}{\partial r}\Big)_{\alpha,l}\ ,\nonumber
\eea
where we introduce the notation
\bea\label{skasisaus}
 {\mathfrak F}=
\frac{r}{2\pi}\  \sin\Big({ \frac{\pi}{2\alpha}}\Big)\  \big(\,  {\mathfrak J}_1+ {\bar   {\mathfrak J}}_1\,\big)\ .
\eea
According to  Ref.\cite{Lukyanov:2010rn} this constant
is  related to the sine-Gordon $k$-vacuum energy $E_k$
\bea\label{sksjsasayusa}
{\mathfrak F}=\frac{R}{\pi}\  (\,E_k-e_\infty\, R\,)\ ,
\eea
provided  $l=2|k|-\frac{1}{2}$, $\alpha=\xi^{-1}$, and
 $e_\infty$  stands for the specific energy  of the system 
with the infinitely large space  size\ \cite{Destri:1990ps}:
\bea\label{vsfd}
e_{\infty}= \lim_{R\to\infty}\frac{E_{k}}{R}= -\frac{ M^2}{4}\ \tan\Big(\frac{\pi\xi}{2}\Big)\ .
\eea
Thus,  for $0<k<\frac{1}{2}$,  relations \eqref{sisissa} are recast into the form
\bea\label{ososaasi}
&& \frac{R}{\pi\xi}\ \Big(\frac{\partial  E_k}{\partial k}\Big)_{r,\xi}=-  
 r\,\Big(\frac{\partial{\hat \eta}_A}{\partial r}\Big)_{\alpha,l}\\
&&\frac{R}{\pi}\  \Big(\frac{\partial  E_k}{\partial \xi}\Big)_{r,k}=-
\frac{r^2}{8\cosh^2(\frac{\pi}{2\alpha})}- {\textstyle \frac{1}{2}}\ 
r\,\Big(\frac{\partial{\Phi }_A}{\partial r}\Big)_{\alpha,l} -l\ r\,\Big(\frac{\partial{\hat \eta}_A}{\partial r}\Big)_{\alpha,l}
\ .\nonumber
\eea

Formulas \eqref{ososaasi} generalize  the FSZ
relations \eqref{sssiasai} and \eqref{siasai} to arbitrary values
of the sine-Gordon coupling constant   and the
quasi-momentum. Indeed,
as  $\xi=2$,
the apex angle of the cone in Fig.\ref{fig1a}a  becomes     $2\pi$, whereas
$k=\frac{1}{4}$ corresponds to $l=0$,  i.e., the  solution of 
the (M)ShG   equation  remains finite at the tip $A$.
In  this special  case,  ${\hat \eta}(w,{\bar w})$ is  expressed in terms  of the  
Painlev${\acute {\rm e}}$ III  transcendent
\eqref{soisiasai},\, \eqref{sosias}:
\bea\label{sososasai}
{\hat \eta}(w,{\bar w})=U\big(\, 4\, |w-w_{B}|\, \big)\ .
\eea
Since $|w_A-w_B|=r/4$ (see Fig.\ref{fig1a}b), the value
${\hat \eta}$
at the apex is given by
\bea\label{soisai}
{\hat {\eta}}_A=  U(r)\ ,
\eea
whereas, as it follows from the general relations $\partial_w\partial_{\bar w}\Phi=4\ \sinh^2({\hat \eta})$ and \eqref{sosssiasai},
\bea\label{uossaisa}
r\, \frac{\rd  \Phi_A}{\rd r}=- \int_r^\infty\rd t\ t\ \sinh^2U(t)\ .
\eea

\subsection{Normalized  on-shell   action }

Although  the  on-shell action  ${\cal A }^*$   disappears from the generalized FSZ relations, it is a main player
in  the   derivation of \eqref{ososaasi}. 
Let us discuss some of its  properties.

The R.H.S. of  \eqref{sksjsasayusa} exponentially decays at $r\to\infty$ (see e.g. \cite{Destri:1994bv}).
This enables us to represent  the  on-shell action in the form
\bea\label{ppsossais}
{\cal  A}^*={\cal A}^*_{\infty}+\int_r^\infty\frac{\rd r}{r}\ {\mathfrak F}\ ,
\eea
where the integration constant    stands for
$\lim_{r\to\infty}{\cal A}^*$. 
The calculations outlined  in Appendix\,\ref{AppendixB} yield its  explicit form
\bea\label{ssisaisa}
{\cal A}^*_\infty&=&
\log\big(3^{\frac{1}{12}} 2^{-\frac{2}{9}}\big)
+(3\xi+1)\ \log\big(A_G\,2^{-\frac{1}{9}}\,\big)\\
&+&2\xi k\
\log\Big(\frac{4k}{\re  }\Big)+
\xi\ \int_{0}^{2k}\rd x\ \log\left(\,\frac{2^{-2 x}\,\Gamma(1- x)}{\Gamma(1+ x)}\,\right)\ ,
\nonumber\eea
where
$A_G=1.28243\ldots$ is Glaisher's constant and we use the sine-Gordon variables $\xi=\alpha^{-1}$ and $k=(2l+1)/4>0$.

The second term in  Eq.\eqref{ppsossais} is of primary interest thus  we  introduce the special  notation 
\bea\label{uayossiasa}
{\mathfrak Y}=\int_R^\infty\frac{\rd R}{\pi}\    (\,E_k-e_\infty\, R\,)\ .
\eea
Evidently it is  the on-shell  ShG action    
normalized by
the condition
\bea\label{jassay}
\lim_{r\to\infty} {\mathfrak Y}=0\ .
\eea
Then Eqs.\eqref{uyososai},\,\eqref{issusossa} are replaced by
\bea\label{sisisaas}
\Big(\frac{\partial {\mathfrak Y}}{\partial k}\Big)_{r,\xi}
&=& 2\xi\ \eta_A-
\Big(\frac{\partial {\cal A}_\infty^*}{\partial k}\Big)_\xi \\
\Big(\frac{\partial {\mathfrak Y}}{\partial \xi}\Big)_{r,k}
&=&{\textstyle \frac{1}{2}}\ \Phi_A+l\,  \eta_A-\Big(\frac{\partial {\cal A}_\infty^*}{\partial \xi}\Big)_k\ , \nonumber
\eea
where we still  assume that  $0<k<\frac{1}{2}$.

Using the  relation  (see the conformal perturbation theory expansion  \eqref{isaiaissus} bellow)
\bea\label{uyast}
\lim_{R\to 0} RE_{k}= -{ \frac{\pi}{6}}\  c_{\rm eff}\ ,
\eea
where
\bea\label{isossiasai}
 c_{\rm eff}=1-\frac{ 24\xi k^2}{1+\xi} 
\eea
is the ``effective'' central charge,
one can represent ${\mathfrak Y}$  in the  form  which is appropriate for the study  of the $R\to 0$  limit,
\bea\label{sisaiaissus}
{\mathfrak Y}= {\textstyle \frac{1}{6}}\ c_{\rm eff}\ \log(MR)+
{\mathfrak Y}_0-\frac{(MR)^2}{8\pi}\ \tan\Big(\frac{\pi\xi}{2}\Big)-
\int_0^R\frac{\rd R}{\pi}\,
\left(\,  E_{k}+ { \frac{\pi  c_{\rm eff}}{6 R}}\, \right)\ .
\eea
Here   $ {\mathfrak Y}_0$ is some $R$-independent  constant.
To calculate this constant explicitly
one should write
${\mathfrak Y}$ as  ${\cal A}^*-{\cal A}^*_{\infty}$,
express  the functional  \eqref{ssiisa} in terms of     the original variables of
the MShG equation\ \eqref{shgz}, and  then analyze the limit of a small $r$.
The straightforward calculations yield
\bea\label{ssisisa}
 {\mathfrak Y}_0 &=&{\textstyle \frac{1}{12}}\ \log\big(\, 4\,
\xi^\xi\,  (1+\xi)^{-1-\xi}\, \big)-
 {\textstyle \frac{1}{6}}\ c_{\rm eff}\ \log(r_\xi)
\\
&-&
\int_0^\infty\frac{\rd x}{x}\
\left(\, \frac{\sinh( x) \cosh(4\xi k x)}{ 2x\sinh(\xi x)\sinh\big(x(1+\xi)\big)
}-\frac{1}{2\xi(1+\xi)\,x^2}+\frac{c_{\rm eff}}{6}\ \re^{-2  x}\, \right)\,  ,\nonumber
\eea
where  $r_\xi$ is given by
Eq.\eqref{tsrars}.

\section{\label{Sect3} YY-function in  the sine-Gordon model}

In this  section  we identify $ {\mathfrak Y}$\ \eqref{uayossiasa} with the  YY-function and
briefly review  the
approach to   numerical calculation 
of its   partial  derivatives.

\subsection{YY-function for  the inhomogeneous  6-vertex model}

\begin{figure}
\centering
\includegraphics[width=10  cm]{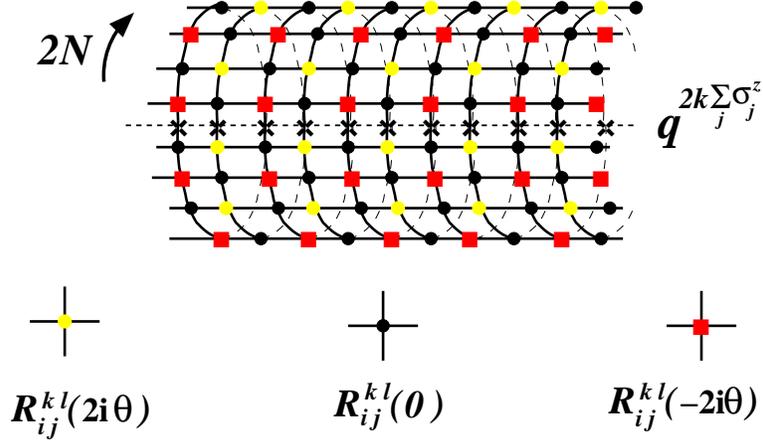}
\caption{ Partition function $Z_N={\rm Tr}\big[\, q^{k\sum_j \sigma_j^z}\, {\boldsymbol \tau}^N\,\big]$
of the  inhomogeneous  6-vertex model on an infinite cylinder. 
Here ${\boldsymbol \tau}$ is the monodromy matrix along the  infinite direction and
$q=\re^{\frac{\ri\pi\xi}{1+\xi}}$. $R_{ij}^{kl}(\lambda)$ are conventional  Boltzmann weights
for the $6$-vertex model satisfying the  Yang-Baxter equation. }
\label{fig5}
\end{figure}

The sine-Gordon model admits an integrable lattice regularization based on
the conventional  $R$-matrix of  the six-vertex model (see Fig.\,\ref{fig5}). 
Here I shall recall   some  basic  facts concerning  the lattice BA equations   which are   relevant  for the purposes of this work.
All the details can be found in Refs.\cite{Destri:1994bv, Destri:1997yz}.

The energy-momentum spectrum in the lattice theory can 
be  calculated by means of the
algebraic BA, or Quantum Inverse Scattering Method: BA state is identified by an unordered  finite set
of distinct, generally complex numbers $\theta_j$ which satisfy  BA equations
\bea\label{iosisisa}
\bigg[\,\frac{s(\theta_j+\Theta+\frac{\ri\pi}{2})
\,s(\theta_j-\Theta+\frac{\ri\pi}{2})}
{s(\theta_j+\Theta-\frac{\ri\pi}{2})
\, s(\theta_j-\Theta-\frac{\ri\pi}{2})}\,\bigg]^N
=-\re^{\frac{4\ri\pi \xi k }{1+\xi}}\ 
\prod_{n}\frac{s(\theta_j-\theta_n+\ri\pi)}
{s(\theta_j-\theta_n-\ri\pi)}\ ,
\eea
where
\bea\label{sossaui}
s(x)=\sinh\Big(\frac{x}{1+\xi}\Big)\ ,
\eea
and $ N$  stands for  one-half of the number    of sites along the compactified  direction in  Fig.\,\ref{fig5}.
The parameter $\Theta$  controls  the  world-sheet inhomogeneity   of  the  Boltzmann weights,
whereas $k$ in \eqref{iosisisa} 
is proportional to the   twist angle for  the quasiperiodic
boundary conditions imposed 
along the compactified  direction. Then
the energy $E^{(N)}$ and   momentum $P^{(N)}$ of the BA state
can be extracted from the formulas
\bea\label{saissais}
\exp\Big(-\ri\  \frac{ E^{(N)}\pm  P^{(N)} }{2N}\,\Big)=
\prod_{j}\frac{s(\frac{\ri\pi}{2}+\Theta\pm\theta_j)}
{s(\frac{\ri\pi}{2}-\Theta\mp\theta_j)}\ .
\eea

For the vacuum state all  the  BA  roots  are real and
their    number  coincides with  $N$,  which is assumed to be even bellow.
Following  Yang and Yang \cite{Yang},  the BA equations in this case
can be bring to the  form of  the  extremum condition
\bea\label{zssisaisa}
\frac{\partial  Y^{(N)} }{\partial \theta_j}=0\ \ \ \ \
\ \ \ \ \
\big(\,  j=-{\textstyle\frac{N}{2}},\, -{\textstyle \frac{N}{2}}+1,\ldots
{\textstyle \frac{N}{2}}-2,\, {\textstyle \frac{N}{2}}-1\,\big) 
\eea
for  the YY-functional defined by the  formulas:
\bea\label{trsossia}
{ Y}^{(N)}=2\ \sum_{j} \Big(\,  V(\theta_j)-\frac{2\xi k\, \theta_j }{1+\xi}\,\Big)
+ \sum_{j,n}\ U(\theta_j-\theta_n)
\eea
with
\bea\label{sosoisai}
V(\theta)=- \frac{N}{\pi}\  \dashint_{-\infty}^\infty
\frac{\rd\omega}{\omega^2}\ 
\frac{
\sinh(\frac{\pi\omega \xi}{2})\, \cos(\omega\Theta)}
{ \sinh(\frac{\pi\omega(1+\xi)}{2})} \ \re^{\ri\omega\theta}
\eea
and
\bea\label{saosisai}
U(\theta)=
\frac{1}{\pi}\  \dashint_{-\infty}^\infty
\frac{\rd\omega}{\omega^2}\ \
\frac{
\sinh(\frac{\pi\omega\xi}{2}) \cosh(\frac{\pi\omega}{2})}
{ \sinh(\frac{\pi\omega(1+\xi)}{2})}\ \re^{\ri\omega\theta}\ .
\eea
Here and bellow the symbol $\dashint$ stands for a   principal value integral 
defined as the half-sum 
 $\frac{1}{2}\, \big(\,
\int_{-\infty-\ri 0}^{+\infty-\ri 0}+
\int_{-\infty+\ri 0}^{+\infty+\ri 0}\, \big)$.

Eqs.\eqref{zssisaisa}    can be interpreted as an  equilibrium condition for  the system of  $N$ 
one-dimensional  ``electrons'' in a  presence
of
confining and linear  external  potentials. 
For large separations, $\theta\gg 1$,
\bea\label{isussoosia}
U(\theta)= -\frac{\xi}{1+\xi}\ |\theta|+O\big(\,\re^{-\frac{2|\theta|}{1+\xi}}\,\big)\ ,
\eea
therefore  the  2-body potential  is essentially   a  1D  repulsive  Coulomb potential
slightly modified at short distances.
Since
\bea\label{soosia}
V(\theta)= \frac{N\xi}{2(1+\xi)}\ |\,\theta-\Theta\,|+
\frac{N\xi}{2(1+\xi)}\ |\,\theta+\Theta\,|+O\big(\,\re^{-\frac{2|\theta\pm \Theta|}{1+\xi}}\,\big)\ ,
\eea
$V(\theta)$
can be interpreted as a  potential produced by
two heavy   positive  charges $+\frac{N\xi}{2(1+\xi)}$ 
placed  at  $\pm \Theta$. We shall always assume
that the  external linear potential in \eqref{trsossia}  is sufficiently  week  and the inequality 
\bea\label{ysoisisai}
-{\textstyle \frac{1}{2}}<k< {\textstyle \frac{1}{2}}
\eea
is fulfilled. For
\bea\label{saosisaui}
0<\xi<1\ ,
\eea
the Hessian of the system \eqref{zssisaisa},
$\frac{\partial^2 Y_N}{\partial \theta_j\partial\theta_n}$, is positive definite therefore 
we will focus primarily on this case.

As the physical intuition suggests, 
the YY-functional \eqref{trsossia}   has   a  stable  minimum at 
some real distribution of the BA roots
\bea\label{sosaosao}
\theta^{(N)}_{-\frac{N}{2}}<\theta^{(N)}_{-\frac{N}{2}+1}<\ldots <\theta^{(N)}_{\frac{N}{2}-2}<\theta^{(N)}_{\frac{N}{2}-1}\ .
\eea
The main subject of our interest is the YY-function, i.e., a critical value of ${ Y}^{(N)}$ calculated at  this minimum.
With  some abuse of notations we will
denote it  by the same symbol as  the  YY-functional, 
\bea\label{sossaisa}
{ Y}^{(N)}={ Y}^{(N)}(\Theta,\,\xi,\, k)\ .
\eea
Using  the YY-function,
the ground state energy \eqref{saissais} can be written as
\bea\label{iasosaisa}
{ E}^{(N)}=\Big(\frac{\partial { Y}^{(N)}}{\partial \Theta}\Big)_{N,\xi,k}\ ,
\eea
whereas
the momentum associated with the  ground state is  of course zero.

\begin{figure}
\centering
\includegraphics[width=9  cm]{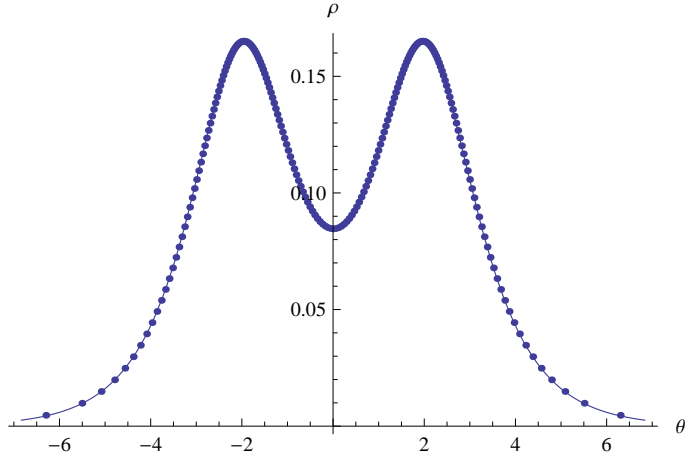}
\caption{$\rho^{(N)}$ from 
Eq.\eqref{bsosiosa} for $N=400$ and $\Theta=2,\, k=0,\, \xi=\frac{2}{3}.$ The solid line represents the
continuous  density \eqref{aiosasau}. }
\label{fig00}
\end{figure}

At large $N$  and finite $\Theta$,
the  distribution of the BA roots
\bea\label{bsosiosa}
 \rho^{(N)}(\theta_{n+\frac{1}{2}}) =\frac{1}{N(\theta_{n+1}-\theta_{n})}\ \ \ \ \ \ \ \ \ 
 \ \ \ \big(\, \theta_{n+\frac{1}{2}}
\equiv {\textstyle \frac{1}{2}}\, (\,
\theta_{n+1}+\theta_{n}\,)\, \big)
\eea
is well approximated by the continuous  density (see Fig.\ref{fig00})
\bea\label{aiosasau}
\rho(\theta)= \frac{1}{2\pi}\ \Big[\, \frac{1}{\cosh(\theta-\Theta)}+ \frac{1}{\cosh(\theta+\Theta)}\, \Big]\ .
\eea

\noindent
Therefore the following limit does  exist
\bea\label{saosoisa}
\lim_{N\to\infty\atop\Theta-{\rm fixed}}N^{-2}\,  { Y}^{(N)}(\Theta)=y_{\infty}(\Theta)\ ,
\eea
and it is a simple  exercise to show that
\bea\label{soissisai}
y_{\infty}(\Theta)=
-\frac{1}{\pi}\ \dashint_{-\infty}^\infty\frac{\rd \omega}{\omega^2}\
\frac{\sinh(\frac{\pi\omega\xi}{2})\ \cos^2(\omega\Theta)}{
\sinh(\frac{\pi\omega(1+\xi)}{2}) \cosh(\frac{\pi\omega}{2})}\ .
\eea
For large $\Theta$,  Eq.\eqref{soissisai} yields
\bea\label{soiass}
y_{\infty}(\Theta)= \frac{\xi\, |\Theta|}{1+\xi}+\frac{y_\infty(0)}{2}+
\frac{2}{\pi}\ \re^{-2|\Theta|}\ \tan\Big(\frac{\pi\xi}{2}\,\Big)+
O\Big(  \re^{-\frac{4|\Theta|}{ 1+\xi}}\Big)\ .
\eea
The constant   term here    coincides  with half of the  value of  \eqref{soissisai} taken at $\Theta=0$.
This is not an  accidental relation. 
Indeed, as   $|\Theta|\to +\infty $, all the BA roots split into  two clusters   centered at 
$\pm \Theta$.
The systems of   BA equations for each   cluster are completely  separated  in this limit
and reduce  to  the original      form  \eqref{iosisisa} with
$\Theta=0$ and $N$ is  replaced by  $N\to N/2$.
Hence for any even  $N$
\bea\label{sikssa}
 { Y}^{(N)}(\Theta)=\frac{\xi N^2}{1+\xi}\ |\Theta|+2\,   Y^{({N}/{2})}(0)+o(1)\ \ \ \  \ \ \ {\rm as}\ \ \ \ \Theta\to\pm \infty\ ,
\eea
where the first term describes   monopole-monopole interaction  of  the ``electron'' clusters
while the second one represents their   intrinsic potential energy.\footnote{In a view of the  mechanical analogy,
it would be  natural  to include an  additional  term
$-\frac{\xi N^2}{1+\xi}\ |\Theta|$ into  the  R.H.S. of  definition\ \eqref{trsossia}.
This term represents the ion-ion potential energy and
does not affect on the equilibrium  conditions\ \eqref{zssisaisa}.}
Combining   \eqref{iasosaisa} with  \eqref{sikssa} one also has
\bea\label{siosisausy}
E^{(N)}(\Theta)- \frac{\xi\, N^2 }{1+\xi}=o(1)\  \ \ \  \ \ \ {\rm as}\ \ \ \ \Theta\to+ \infty\ .
\eea
It should be emphasized that  asymptotic formulas  \eqref{sikssa} and \eqref{siosisausy}
do not assume the  large-$N$ limit and can be applied   for any finite $N$.

\subsection{Scaling limit}

The sine-Gordon   QFT \eqref{sg} manifests itself in
the
scaling limit when both $N,\, \Theta\to+\infty$ while
the scaling  parameter
\bea\label{sissa}
r=4\, N\ \re^{-\Theta}
\eea
is kept fixed  (RG-invariant).
In this case   the L.H.S. of \eqref{siosisausy}  does not vanish, but
has a simple relation to the  $k$-vacuum energy   \cite{Destri:1994bv}:
\bea\label{gsossia}
\lim_{N,\Theta\to+\infty\atop
r-{\rm fixed}}\ \left(\, { E}^{(N)}-\frac{\xi\, N^2}{1+\xi}\ \,\right)=
 \frac{RE_k }{2\pi}+  \frac{ c_{\rm eff}}{12}\ ,
\eea
where the effective center charge $c_{\rm eff}$ is given by  Eq.\eqref{isossiasai}.

In order to study the scaling behavior of  the YY-function,  it makes sense
to consider only the  part of the  ``electron-ion'' potential energy   corresponding to   the mutual  interaction of
the  clusters,
\bea\label{sossisa}
{ Y}_{\rm int}^{(N)}(\Theta)= { Y}^{(N)}(\Theta)-\frac{\xi N^2}{1+\xi}\ |\Theta|-2\,   Y^{(N/2)}(0)\ ,
\eea
which   vanishes as $\Theta\to+\infty$ for any fixed $N$. 
Taking into account Eqs.\eqref{iasosaisa},\,\eqref{gsossia}, we get
\bea\label{issai}
\lim_{N, \Theta\to+\infty\atop
r-{\rm fixed}} { Y}_{\rm int}^{(N)}(\Theta)=-\int_0^r\frac{\rd r}{\pi r}\ 
\left(\, RE_k+ { \frac{\pi  c_{\rm eff}}{6 }}\, \right)\  ,
\eea
or, equivalently using \eqref{sisaiaissus},\footnote{
Notice that  $Y^{(N)}(0)$ can be interpreted as  the YY-function for  the  spin-$\frac{1}{2}$ Heisenberg chain.
As $N\to\infty$ $$Y^{(N)}(0)=  y_{\infty}(0)\ N^2+
 {\textstyle \frac{1}{6}}\ c_{\rm eff}\   \log\big({\textstyle \frac{\pi N }{4}}\big)+
{\mathfrak Y}_0+o(1)\, ,$$
where ${\mathfrak Y}_0$ and  $y_\infty(0)$   are  given by Eqs.\eqref{ssisisa} and \eqref{soissisai}, respectively.}
\bea\label{issaiioi}
\lim_{N, \Theta\to+\infty\atop
r-{\rm fixed}} { Y}_{\rm int}^{(N)}(\Theta)=
{\mathfrak Y}- {\textstyle \frac{1}{6}}\ c_{\rm eff}\ \log(r)+\frac{r^2}{8\pi}\ \tan\Big(\frac{\pi\xi}{2}\Big)-{\mathfrak Y}_0\ .
\eea
The last formula allows one  to identify the
YY-function in the sine-Gordon QFT with the normalized on-shell action ${\mathfrak Y}$.

The following comment is in order here.
Our analysis is  based on the existence of solutions \eqref{sosaosao}
of the vacuum BA equations.  It
can be directly applied to
the case $0<\xi<1$ only.  At   $\xi=1$, the sine-Gordon model is equivalent to the  free massive Dirac fermions theory
and  a closed form of  the  YY-function can be easily derived  from  definition\ \eqref{uayossiasa}:
\bea\label{sssiossisa}
{\mathfrak Y}=
-r\ \int_{-\infty}^\infty\frac{\rd\tau}{ 2\pi^2}\ \tau\, \sinh(\tau)\, \log\Big[\,\big(1+\re^{-r\cosh(\tau)+2\pi \ri  k})
(1+\re^{-r\cosh(\tau)-2\pi \ri k}\big)\, \Big]\ .
\eea
It is expected that  for $\xi>1$   the YY-function      is   uniquely defined
through the analytic continuation from the segment $\xi\in (0,1)$.

\subsection{BA roots at  the large-$N$ limit}

Properties of the BA roots at the scaling limit  were  discussed    in  Ref.\,\cite{Destri:1997yz}.
The roots accumulate   at  $\theta=\pm \Theta$.  However,
at  the   center region (see Fig.\ref{fig00}) and at   the tails of the distribution
the  roots remain isolated and
their behavior can be     described as follows
(see Tables\,\ref{F-Cyns} and \ref{Fa-Cyns} for  illustration):

\begin{table}
\begin{center}
\begin{tabular}{| c || l | l | l | l | l |}
\hline \rule{0mm}{3.6mm}
$n$ & $N=100$ & $N=200$ & $N=400$ & $N=800$ & $N=1600$ \\
\hline
$0$  & 1.04348   & 1.04342    &1.04340   &1.04340 & 1.04340   \\
$1$  & 3.01807   & 3.01640    &3.01598   &3.01588 & 3.01585   \\
$2$  & 5.01975   & 5.01202    &5.01009   &5.00960 & 5.00948  \\
$3$  & 7.03507   & 7.01378    &7.00848   &7.00716 & 7.00683  \\
$4$  & 9.06565   & 9.02023    &9.00896   &9.00615 & 9.00545  \\
$5$  & 11.1150   & 11.0317    &11.0111   &11.0059 & 11.0046  \\
$6$  & 13.1873   & 13.0489    &13.0149   &13.0064 & 13.0043  \\
$7$  & 15.2870   & 15.0729    &15.0204   &15.0074 & 15.0042 \\
$8$  & 17.4186   & 17.1044    &17.0280   &17.0090 & 17.0043  \\
$9$  & 19.5874   & 19.1447    &19.0378   &19.0112 & 19.0046  \\
\hline
\end{tabular}
\end{center}
\caption{ $\frac{r}{2\pi}\  \exp\big(\theta^{(N)}_n\big)$  for $r=1$, $\xi=\frac{2}{3}$ and $k=0$}
\label{F-Cyns}
\end{table}

\begin{table}
\begin{center}
\begin{tabular}{| c || l | l | l | l | l |}
\hline \rule{0mm}{3.6mm}
$n$ & $N=100$ & $N=200$ & $N=400$ & $N=800$ & $N=1600$ \\
\hline
$0$  &  1.02837   & 1.02831    &1.02830   &1.02830 &  1.028299    \\
$1$  &  3.01276   & 3.01111    &3.01069   &3.01059 &  3.01056   \\
$2$  &  5.01652   & 5.00881    &5.00689   &5.00641 &  5.00628  \\
$3$  &  7.03273   & 7.01148    &7.00619   &7.00487 &  7.00454  \\
$4$  &  9.06381   & 9.01843    &9.00718   &9.00436 &  9.00366  \\
$5$  &  11.1135   & 11.0302    &11.0096   &11.0045 &  11.0032  \\
$6$  &  13.1860   & 13.0477    &13.0136   &13.0051 &  13.0030  \\
$7$  &  15.2858   & 15.0718    &15.0194   &15.0063 &  15.0031 \\
$8$  &  17.4176   & 17.1035    &17.0271   &17.0081 &  17.0033  \\
$9$  &  19.5864   & 19.1438    &19.0369   &19.0104 &  19.0038  \\
\hline
\end{tabular}
\end{center}
\caption{ $\frac{4 N}{\pi}\  \exp\big(\theta^{(N)}_{n-\frac{N}{2}}+\Theta\big)$  for $r=1$, $\xi=\frac{2}{3}$ and $k=0$}
\label{Fa-Cyns}
\end{table}

\begin{itemize}

\item
There exist  limits
\bea\label{xsosasai}
\theta_j=\lim_{N, |\Theta|\to\infty\atop
r,\, j-{\rm fixed} }\theta_j^{(N)}\  \ \ \ \ \  \ \ (\, j=0,\,\pm1\,\pm 2\ldots\, )
\eea
and,  for an arbitrary  $\Theta\geq 0$,
\bea\label{sisusossais}
{\tau}^{(+)}_n&=&\lim_{N\to\infty\atop
n-{\rm fixed}} \Big(\,  \theta^{(N)}_{n-\frac{N}{2}}+\log(N)+\Theta\, \Big)
 \\
{\tau }^{(-)}_n&=& \lim_{N\to\infty\atop
m-{\rm fixed}}\Big(\, \theta^{(N)}_{\frac{N}{2}-1-n} -\log(N)-\Theta\,\Big)
\ \  \ \ \ \ \ \ \ \ \ \ (\, n=0, \,1\ldots\, )  \ .\nonumber
\eea

\item
The limiting values of the roots 
possess the following   $n\to+\infty$  asymptotic behavior
\bea\label{saklsisais}
\re^{\theta_n}&=&
\frac{2\pi}{r}\ \big(\, 2n+1+
2k\,\big)+
O\big(n^{-1}\big)\\
\re^{-\theta_{-n-1}}&=&
\frac{2\pi}{r}\ \big(\, 2n+1-
2k\,\big)+O\big(n^{-1}\big)
 \nonumber
\eea
and
\bea\label{ossuusiosisa}
\exp\big(\pm \tau^{(\pm)}_n\big)=\frac{\pi}{2}\ 
\big(\, 2n+1\pm 2 k\,\big)+O\big(n^{-1}\big)\ .
\eea

\end{itemize}

\noindent
To probe  
the  infinite sequences
$\{\theta_j\}_{j=-\infty}^\infty$ and 
$\{\tau^{(\pm)}_n\}_{n=0}^\infty$  
it is useful to consider
certain generating functions encoding  their properties.
Let $\zeta_+(\omega)$  and  $\zeta_-(\omega)$ be  functions defined  as  the analytic continuation of
convergent series
\bea\label{ssisai}
\zeta_+(\omega)&=& \sum_{n=0}^\infty\re^{-\ri\omega\theta_n}\ \ \ \ \ \ \ \ \ \ \ \Im m (\omega)<-1\ ,\nonumber\\
\zeta_-(\omega)&=& \sum_{n=0}^\infty\re^{-\ri\omega\theta_{-1-n}}\ \ \ \ \ \ \ \,\Im m(\omega)>1\ .
\eea
As follows from  the asymptotic formulas  \eqref{saklsisais},
$\zeta_+(\omega)\ \big(\,\zeta_-(\omega)\,\big)$  is  analytic in the half plane $\Im m(\omega)< 1$  $(\Im  m (\omega)>- 1)$ except at
a simple pole at $ \omega=-\ri$  $(\omega=\ri)$ with the residue $ - \frac{\ri r}{4\pi  }$  $ \big( \frac{\ri r}{4\pi  }\big)$.
Also note that   $ \zeta_\pm(0)=\mp k$.
Therefore
\bea\label{sososasau}
\zeta(\omega)=\zeta_+(\omega)+\zeta_-(\omega)\ ,
\eea
is an analytic function in the strip $|\Im m(\omega)|<1$ such that
\bea\label{sususay}
 \zeta(0)=0\ .
\eea
In the limit $r\to 0$
\bea\label{oaasauas}
\zeta(\omega)= \Big(\frac{r}{4}\Big)^{\ri\omega}\ \big(\,\zeta^{\rm(cft)}_k(\omega)+o(1)\,\big)+
\Big(\frac{r}{4}\Big)^{-\ri\omega}\ \big(\, \zeta^{\rm(cft)}_{-k}(-\omega)+o(1)\,\big)\ ,
\eea
where $\zeta^{\rm(cft)}_{\pm k}(\pm \omega)$
are the  zeta  functions  for the sequences $\{\tau^{(\pm )}_n\}_{n=0}^\infty$ \eqref{sisusossais}, i.e.,
\bea\label{sisasaaus}
\zeta^{\rm(cft)}_{\pm k}(\pm \omega)=\sum_{n=0}^\infty \exp\big( - \ri\omega\tau^{(\pm)}_n\,\big)\ .
\eea
The  function\ $\zeta^{\rm(cft)}_{ k}(\omega)$ was introduced (in a different overall normalization) and
studied in Ref.\cite{Bazhanov:1996dr}.
Its  exhaustive   description 
was found later   in Ref.\cite{Bazhanov:1998wj} (see also related works \cite{Voros:1992}
and \cite{Dorey:1998pt}), where
it was shown that  $\zeta^{\rm(cft)}_{ k}(\omega)$  coincides with the zeta function of  
the  Schr${\ddot {\rm o}}$dinger operator
\bea\label{xssiisa}
-\partial_x^2+ x^{2\alpha}+\frac{l(l+1)}{x^2}\ .
\eea
More precisely,  for $0< k <\frac{1}{2}$,
the sequences  $\big\{{\tau}^{(\pm)}_n\,\big\}_{n=0}^\infty$
are simply related to
the spectral sets  $ \big\{{\cal E}^{(\pm)}_n\}_{n=0}^\infty$ of  this differential operator:
\bea\label{saskalska}
\exp\big(\pm {\tau}^{(\pm)}_n\big)=\frac{\xi r_\xi}{8}\ \
\left(\, {\cal E}_n^{(\pm)}\, \right)^{\frac{1+\xi}{2}}\ ,
\eea
with              
$\alpha=\xi^{-1}, \, l=2k-\frac{1}{2}$ and
$r_\xi$  given by \eqref{tsrars}.

With the above properties of the BA roots it is not  difficult to analyze
the large $N$-limit of the relation
\bea\label{uspssa}
\Big(\frac{\partial Y^{(N)}}{\partial k}\Big)_{N,\Theta,\xi}=-\frac{4\,\xi }{1+\xi}\ \sum_j\theta^{(N)}_j\ ,
\eea
which is derived by differentiating   \eqref{trsossia}
with the use of
BA equations \eqref{zssisaisa}.  The scaling analog of \eqref{uspssa}
reads
\bea\label{spssyssta}
\Big(\frac{\partial {\mathfrak Y}}{\partial k}\Big)_{r,\xi}=-\frac{4\ri\,\xi }{1+\xi}\ { \zeta}'(0)\ ,
\eea
where prime stands for the derivative with respect to  $\omega$. As
follows from\ \eqref{oaasauas} and $\zeta^{\rm(cft)}_k(0)=-k$,
\bea\label{siadsayqq}
\Big(\frac{\partial {\mathfrak Y}}{\partial k}\Big)_{r,\xi}=
-\frac{8k\,\xi }{1+\xi}\  \log\Big(\frac{r}{4}\Big)
-\frac{4\ri\,\xi }{1+\xi}\ \left(\, \partial_\omega\zeta^{\rm(cft)}_k(0)+
\partial_\omega\zeta^{\rm(cft)}_{-k}(0)\, \right) +o(1)\ \ \ \ \ {\rm as}\ \ \ r\to 0\ .
\eea
The subleading term of  this asymptotic is expressed in terms of the determinant of the differential
operator \eqref{xssiisa} and can be calculated explicitly:
\bea\label{sossias}
\Big(\frac{\partial{\mathfrak Y}}{\partial p}\Big)_{r,\xi}=
\log\bigg[\,  \Big(\frac{r}{r_\xi}\Big)^{-\frac{2p}{\xi(1+\xi)}}\ 
\xi^{\frac{2p}{\xi}}\
(1+\xi)^{-\frac{2 p}{1+\xi}}\ \
\frac{\Gamma(1+\frac{p}{\xi})\Gamma(1-\frac{p}{1+\xi})}{\Gamma(1-\frac{p}{\xi})
\Gamma(1+\frac{p}{1+\xi})}\, \bigg]+o(1)\    \ {\rm as}  \ \ r\to 0\, ,
\eea
where we substitute  $k$ for   the equivalent parameter  $p=2\xi k$.
Of course, Eq.\eqref{sossias} can be alternatively obtained  by means of relations \eqref{sisaiaissus},\,\eqref{ssisisa}.
Note  the expression in the bracket $\big[\,\cdots\,\big]$ coincides with
Liouville reflection amplitude (analytically continued to  the domain $-1<\xi<0$)  
introduced in the work  \cite{Zamolodchikov:1995aa}.

\subsection{Calculation of partial derivatives of the YY-function}

For $0<\xi<1$ the so-called $Q$-function can be defined through  the  convergent product
\bea\label{saossisa}
Q(\theta)={\mathfrak C}\ \ \re^{\frac{2k \theta}{1+\xi}}\ \prod_{n=0}^\infty
4\ \re^{\frac{\theta_{-n-1}-\theta_n}{1+\xi}}\
s\big(\theta_n-\theta\big)\, s\big(\theta- \theta_{-n-1}\big)\ ,
\eea 
where the  abbreviation $s(\theta)$
\eqref{sossaui} is applied.
The  $\theta$-independent factor    ${\mathfrak C}$ can be chosen at will.
In what follows it is assumed that
\bea\label{skskssisai}
{\mathfrak C}= 2^{\frac{1}{1+\xi}}\
\prod_{n=0}^{\infty}\ \exp\Big[ \,
{\textstyle \frac{1}{1+\xi}}\ \big(\,
 2\, \log\big({\textstyle \frac{r}{2\pi (2n+1)}}\big)
+\theta_n- \theta_{-n-1}\,\big)
\, \Big]\ .
\eea
In the scaling limit,  BA equations  \eqref{zssisaisa}
boil down to
\bea\label{soisisa}
\epsilon(\theta_j)=\pi\, (2j+1)\ \ \ \ \ \ (j=0,\,\pm 1,\,\pm 2\ldots\, )\ ,
\eea
where
\bea\label{sosoia}
\epsilon(\theta)=\ri \log\left(\,\frac{Q(\theta+\ri\pi\xi)}{Q(\theta-\ri\pi\xi)}\,\right)
\eea
and the branch of the log is fixed by the condition
\bea\label{ssosisasiu}
\epsilon(\theta)= \frac{r\re^\theta}{2}-2\pi k+o(1)\ \ \ \ \ \ \ \ \ \ {\rm for}\ \ \ \ \ \  \Re e(\theta)\to+\infty \ \ \ \ 
{\rm and}\ \ \ |\Im m (\theta)|<\frac{\pi}{2}\ .
\eea

Using the analytic properties of $\zeta_\pm(\omega)$ \eqref{ssisai},
definitions \eqref{saossisa} and \eqref{sosoia}  can be transformed  
into the integral representations
\bea\label{jsaospsosa}
\log Q\big(\theta+{\textstyle \frac{\ri\pi (1+\xi)}{ 2}} \big)=\ri\pi k+
\frac{r\cosh(\theta)}{2 \cos(\frac{\pi\xi}{ 2}) }-
\frac{1}{2}\ \dashint_{-\infty}^\infty\frac{\rd\omega}{\omega }\
\frac{{ \zeta}(\omega)}{\sinh(\frac{\pi\omega(1+\xi)}{2})}\ \re^{\ri\omega\theta}
\eea
and
\bea\label{soisoai}
\epsilon(\theta)=-2\pi k 
+r\ \sinh(\theta)-\ri\ \int_{-\infty}^\infty\frac{\rd\omega}{\omega }\
\frac{ \sinh(\frac{\pi\omega(1-\xi)}{2}) }{\sinh(\frac{\pi\omega(1+\xi)}{2})}\  \zeta(\omega)\ \re^{\ri\omega\theta}\ ,
\eea
respectively.
On the other hand,  the BA equations\ \eqref{soisisa} imply
(see Ref.\cite{Destri:1992qk} and related Ref.\cite{Klumper:1991}  for the original derivation)
\bea\label{sksklsskla}
\zeta(\omega)=
\frac{\ri\, \omega\, \sinh(\frac{\pi\omega (1+\xi)}{ 2})}{
 \cosh(\frac{\pi\omega}{ 2})\,
\sinh(\frac{\pi\xi \omega}{ 2})}\
\int_{-\infty}^{\infty}  \frac{\rd\theta}{2\pi}\  \re^{-\ri\omega \theta}\,
\Im
m\Big[\,   \log\big(1+\re^{-\ri \epsilon(\theta-\ri 0)}\, \big)
\,\Big]\ .
\eea
Note that  at the free-fermion point  ($\xi=1$) $\epsilon(\theta)=r\, \sinh(\theta)-2\pi k$,
therefore Eq.\eqref{sksklsskla} gives
\bea\label{ssaisisai}
\zeta(\omega)= \omega \int_{-\infty}^\infty\frac{\rd\theta}{2\pi}\, \re^{-\ri\omega\theta}
\Big[\, \re^{-\frac{\pi\omega}{2}}\ \log\big(1+\re^{-r\cosh(\theta)+ 2\pi\ri  k}\,\big)-
\re^{\frac{\pi\omega}{2}}\ \log\big(1+\re^{-r\cosh(\theta)- 2\pi\ri  k}\,\big) \Big] .
\eea
In general,
Eqs.\eqref{soisoai} and \eqref{sksklsskla} are combined  into a single integral equation on  $\epsilon(\theta)$.
Once the  numerical data  for  $\epsilon(\theta)$ are available,  
$\zeta(\omega)$ can be computed by means of  \eqref{sksklsskla}.

Eq.\eqref{sksklsskla} shows that $\zeta(\omega)$
is a meromorphic function  with  simple poles located at 
$\omega=\pm \ri\, (2 n+1)$ and $\omega= \pm \frac{2\ri}{\xi}\  (n+1) \ (n=0,\,1\ldots)$.
The residue values are
the  $k$-vacuum eigenvalues of local and nonlocal integral of motions in the
quantum sine-Gordon model\ \cite{Bazhanov:1996dr,Lukyanov:2010rn}.
In particular, at the boundary of the strip  of analyticity  $|\Im m(\omega)|<1$,  $ \zeta(\omega)$  has simple poles 
\bea\label{uysoosaao}
 \zeta(\omega)=  \mp 
 \frac{\mathfrak F}{r} \cot\Big(\frac{\pi\xi}{2}\Big)\   \frac{\ri  }{  \omega\pm \ri }+O(1)\ \ \ \ \ {\rm as}\ \ \ \ 
\omega\to\pm\ri\ ,
\eea
where ${\mathfrak F}$ is given by Eq.\eqref{sksjsasayusa}.

This way,  the   problem of numerical   calculation
of  $k$- and $r$-partial derivatives of the YY-function can be solved by means of relations
\bea\label{usystsoissisa}
\Big(\frac{\partial{\mathfrak Y}}{\partial r}\Big)_{\xi,k}
&=& \tan\Big(\frac{\pi\xi}{2}\Big)\ \lim_{\omega\to\mp\ri} (1 \mp\ri\, \omega)\,\zeta(\omega)\ ,\nonumber \\
\Big(\frac{\partial{\mathfrak Y}}{\partial k}\Big)_{r,\xi}&=&-\frac{4\ri\, \xi   }{1+\xi}\ \zeta'(0)\ .
\eea
The calculation of $\xi$-derivative is found out  to be a more delicate problem.
Rather naive manipulations with  the lattice  YY-functional\ \eqref{trsossia} suggest
that
\bea\label{assjsajasju}
\Big(\frac{\partial{\mathfrak Y}}{\partial \xi}\Big)_{r, k}&=&
\frac{1}{4}\ \int_{-\infty}^\infty
\frac{\rd\omega}{\omega}\
\frac{ \sinh(\pi \omega)}
{ \sinh^2(\frac{\pi\omega(1+\xi)}{2})}\ \zeta(\omega)\,
\zeta( -\omega)\ .
\eea
In Appendix\,\ref{AppendixC}
we present some evidences in  support of this relation.
Unfortunately    it  still lacks a rigorous proof.

\section{\label{SectionMini} Minisuperspace  limit}

\subsection{Minisuperspace limit   of the YY-function}

The small-$R$ expansion of the $k$-vacuum   energy in the quantum  sine-Gordon model
was argued  in  Ref.\cite{Zamolodchikov:1994uw},
\bea\label{isaiaissus}
\frac{RE_k}{\pi}=-  { \frac{1}{6}}+\frac{4k^2\xi}{1+\xi  }-
\sum_{n=1}^\infty\,   f _n\  r^{\frac{4  n}{1+\xi}}\ .
\eea
The  first    coefficient $f_1$ has a relatively simple explicit form
\bea\label{soisasa}
  f_1= 4\  \ 
\frac{ \gamma^2(\frac{\xi}{1+\xi})\gamma(\frac{\xi(1- 2k)}{1+\xi})\gamma(\frac{\xi(1+2k)}{1+\xi})}{\gamma(\frac{2\xi}{1+\xi})
\,  (\, (1+\xi)\, r_\xi\,)^{{\frac{4}{1+\xi}}}}\ ,
\eea
where $\gamma(x)=\Gamma(x)/\Gamma(1-x)$ and  $r_\xi$ \eqref{tsrars}.
Let us consider
the  $\xi\to 0$  limit of \eqref{isaiaissus} 
in which   the  parameters $r$ and  $k$
are kept   fixed. One finds
\bea\label{ssasoaaiso}
\frac{RE_k}{\pi}=-\frac{ 1}{6}+ \xi\ a+o(\xi)
\eea
with
\bea\label{sisaasu}
a=\nu^2+\frac{q^2}{2\,(\nu^2-1)} +O(q^4)\ .
\eea
Here  we have denoted
\bea\label{usoisisai}
q =\Big(\frac{r}{4}\Big)^2\ ,\ \ \ \ \ \ \ 
\nu=2 k\ .
\eea
In this special (minisuperspace) limit 
the sine-Gordon QFT reduces to the quantum mechanical problem  of
particle in the cosine potential
whose  energy  coincides with
$a$ from Eq.\eqref{ssasoaaiso}.\footnote{
The minisuperspace approximation for the closely related quantum sinh-Gordon  model  was discussed in Ref.\cite{Lukyanov:2000jp}.}
Note that \eqref{usoisisai} are conventional notations  in the theory of Mathieu
equation  \cite{Stegun}. 
For given $\nu$,  $a$  is determined  by the
Whittaker equation
\bea\label{sakssaisa}
\sin^2\Big(\frac{\pi\nu}{2}\Big)=\Delta_q( a)\ \sin^2 \Big(\frac{\pi\sqrt{a}}{2}\Big)\ ,
\eea
where $\Delta_q(a)$ is   Hill's determinant
\bea\label{salososa}
\Delta_q(a)=
\det \begin{pmatrix}
&\cdots&  \cdots     &   \cdots       &  \cdots         &  \cdots       & \cdots  & \cdots&     \\
&\cdots&   \gamma_{-2}& 1        & \gamma_2 & \cdots        &  \cdots   & \cdots &    \\
&\cdots&  \cdots            &\gamma_{0}& 1        &\gamma_0 & \cdots  & \cdots &     \\
&\cdots& \cdots              &  \cdots        & \gamma_2 & 1       & \gamma_2 & \cdots & \\
&\cdots&  \cdots         &   \cdots       &  \cdots        &   \cdots      & \cdots & \cdots &
\end{pmatrix}\ ,\ \ \ \ \ \ \ \  \gamma_{2n}=\frac{q}{4n^2-a}\  .
\eea
The solution
of  Eq.\eqref{sakssaisa} is a  multivalued function, but  the condition \eqref{sisaasu}
specifies the proper  branch  unambiguously.\footnote{It 
is   implemented in  the $Mathematica$ as
 ${\rm MathieuCharacteristicA}[\,\nu,\, q\,]$ with $-1<\nu<1$.}
To simplify formulas,   we will below     treat  $a$   as a function of the variables $r$ and $\nu$, i.e.,
\bea\label{aauaussay}
a=a(r,\nu)\ .
\eea

Now, using Eqs.\eqref{sisaiaissus} and \eqref{ssasoaaiso}  it is  straightforward to
obtain the limiting behavior of  the YY-function:
\bea\label{osaosasao}
{\mathfrak Y}=
\frac{1}{6}\ \log\left(\frac{ r\xi   A_G^{12}}{4}\right)+
\xi\ \left(\, \frac{1}{4}\, 
\log\big(\, \re^{-\frac{1}{3}}\, 2^{-\frac{2}{3}}\,\xi\,\big)+ {\cal Y}(r,\nu)\, \right)+o(\xi)\ ,
\eea
where
\bea\label{soisosa}
{\cal Y} (r,\nu)&=&{\cal Y}_0(\nu)
-\nu^2\ \log\Big(\frac{r}{4}\Big)-\frac{r^2}{16}
-\int_0^{r}\frac{\rd t}{t}\ \big(\, a(t,\nu)-\nu^2\,\big)\ ,
\eea
with
\bea\label{tskssia}
{\cal Y}_0(\nu)=
\int_0^\infty\frac{\rd x}{x}\
\left(\, \nu^2-\frac{ \sinh^2(\nu x)}{x\sinh(x)}
 \, \right)\ \re^{-x} \ .
\eea
Note that
the above form  of   ${\cal Y}( r,\nu)$ can be alternatively    rewritten as\footnote{ For  the large-$r$ expansion of $a$
see, e.g., formulas 28.8.1, 28.8.2 in Ref.\cite{Wolf}.}
\bea\label{sioisai}
{\cal Y}( r,\nu)=
 \log\big(2^{\frac{1}{6}}\, A_G^3\big)+
\frac{1}{4}\ \log(r)-\frac{r}{2}+
\int_{r}^\infty\frac{\rd t}{ t}
\ \Big(\, a(t,\nu)+\frac{t^2}{8}-\frac{t}{2}+\frac{1}{4}\, \Big)\ .
\eea

\subsection{\label{MiniQ}Minisuperspace limit  of  the $Q$-function} 

The minisuperspace approximation of the $Q$-function at  $r=0$   was argued in Appendix B
of Ref.\cite{Bazhanov:1996dr}.
The analysis was  based on    general 
properties of the $Q$-function and
only minor adjustments are needed to extend it  to the case $r>0$.

The  $Q$-function   is a quasiperiodic solution  (see Eq.\eqref{saossisa})
\bea\label{iskissasua}
Q\big(\theta+\ri\pi(1+\xi)\, \big)=\re^{2\ri\pi k}\ Q(\theta)
\eea
of Baxter's equation  (see,  e.g., Ref.\cite{Lukyanov:2010rn})
\bea\label{isssisa}
T_{\frac{1}{2}}(\theta)\, Q(\theta)=Q\big(\,\theta+\ri\pi\xi\, \big)+
Q\big(\,\theta-\ri\pi\xi\, \big)\ ,
\eea
where  $ T_{\frac{1}{2}}(\theta)$ stands for the  $k$-vacuum eigenvalue
of the  transfer matrix.  If the 
overall normalization factor   in \eqref{saossisa} 
is chosen as in Eq.\eqref{skskssisai}, then the $Q$-function
also 
obeys the so-called quantum Wronskian relation
\bea\label{hasta}
Q\big(\theta+{\textstyle\frac{\ri\xi\pi}{2}}, \, k\, \big)\,
Q\big(\theta-{\textstyle\frac{\ri\xi\pi}{2}}, -\, k\, \big) -
Q\big(\theta-{\textstyle\frac{\ri\xi\pi}{2}}, \,k \, \big)\,
Q\big(\theta+{\textstyle\frac{\ri\xi\pi}{2}}, -\, k\, \big)=2\ri\sin(2\pi k )\ ,
\eea
where we explicitly indicate the dependence on  the quasi-momentum.
In the minisuperspace limit
\bea\label{yiusosisai}
T_{\frac{1}{2}}(\theta)=2+(\pi\xi )^2\ w(\theta)+o(\xi)\ ,
\eea
whereas Baxter's  equation reduces to the second order differential equation.
For the conformal case (i.e., for  $r=0$) discussed in \cite{Bazhanov:1996dr}, 
 $w(\theta)=\re^{2\theta}-(2k)^2$. Similarly, in the case of finite $r$  one can show that
\bea\label{jaausy}
w(\theta)=\frac{r^2}{8}\ \cosh(2\theta)-A\ ,
\eea
 with some $\theta$-independent constant $A$ such that 
\bea\label{kausya}
\lim_{r\to 0}\, A=(2k)^2\ .
\eea
Thus the minisuperspace limit  of the  $Q$-function can be described as follows.
Let $F_\nu(z)$ be 
Floquet's solution 
\bea\label{issuas}
F_\nu(z+\ri\pi)=\re^{\ri \pi\nu}\ F_\nu(z)\ ,\ \ \ \ \ \ \ \ F_{-\nu}(z)=F_{\nu}(-z)
\eea
of the modified 
Mathieu equation
\bea\label{yqtqrs}
- \frac{\rd^2 F}{\rd z^2}+\Big(\, A-\frac{r^2}{8} \, \cosh(2z )\,\Big)\, F=0
\eea
normalized by the condition
\bea\label{iosausau}
W[F_\nu, F_{-\nu}]=-2\ \sin(\pi\nu )\ ,
\eea
where 
$W[f,g]$ stands for  the Wronskian $f g'-g f'$.
Then, as follows from \eqref{iskissasua}-\eqref{jaausy},
\bea\label{aosasu}
\lim_{\xi\to 0\atop
\theta,\,r,\, k-{\rm fixed}}\, \sqrt{\pi \xi}\ Q(\theta, k)=  F_\nu(\theta)\ \ \ \ \ \ \ \ \ \ {\rm with}\ \ \ \ \ \  \ \ \nu=2k\ .
\eea
For given $\nu$, the constant $A$ in  \eqref{yqtqrs} is determined by
the quasiperiodicity condition \eqref{issuas}, which implies
the Whittaker equation  \eqref{sakssaisa} with     $a$   replaced by $A$. 
The extra condition  \eqref{kausya} enables us to chose
the  branch of  solution of \eqref{sakssaisa} unambiguously.
Thus we   conclude that 
\bea\label{sosasai}
A=a(r,\nu)\ ,
\eea
where $a$ is the same function   as in Eqs.\eqref{ssasoaaiso},\,\eqref{aauaussay}.
It is useful to  note that Eq.\eqref{iosausau}
is  equivalent to the following  normalization 
condition\footnote{ For $0< \nu<1$$,
F_\nu(z)={\cal N}_\nu\, \big(\,ce_\nu(\ri\, z)-\ri\, se_\nu(\ri\, z)\, \big)\,,\ \ {\cal N}_\nu=
\sqrt{\frac{\sin(\pi\nu)}{se'(0)ce_\nu(0)}}$, where     $ce_\nu( x)$ and $ se_\nu(x)$
are returned by the
$Mathematica$ functions ${\rm MathieuC}[a,\, q,\, x]$ and ${\rm MathieuS}[a,\, q,\, x]$, respectively.
Their $x$-derivatives are implemented as
${\rm MathieuCPrime}[a,\, q,\, x]$ and ${\rm MathieuSPrime}[a,\, q,\, x]$.  Here  $a=a(r,\nu)$ and
$q=\big(\frac{r}{4}\big)^2$.}
\bea\label{ksaskla}
\int_0^\pi \rd y\ F_\nu (\ri\, y)\, F_\nu (-\ri\, y)=2\pi\sin(\pi \nu)\ \bigg[\, 
\Big(\frac{\partial a}
{\partial \nu}\Big)_r\, \bigg]^{-1}\ .
\eea

Eq.\eqref{aosasu} dictates  that the BA roots $\{\theta_j\}_{j=-\infty}^\infty$   \eqref{xsosasai}
in the minisuperspace limit turn out  to be the  zeros of the  Mathieu
function,\footnote{The remaining  terms $O(n^{-1})$ in  the  large-$n$ asymptotic 
formulas  \eqref{saklsisais} diverge as $\xi\to 0$.
For this reason these   formulas are not applicable at
the minisuperspace limit.}
\bea\label{ysasaioaois}
\lim_{\xi\to 0\atop j.\,r,\,k-{\rm fixed}}\theta_j=z_j\ \ \ :\ \ \ F_\nu(z_j)=0\ \ \ \ \ (j=0,\,\pm1,\pm2\ldots\,)\ .
\eea
Using the  properties of  $F_\nu(z)$, it is not difficult to derive the  sum rule
\bea\label{sskssia}
\Sigma=\sin(\pi\nu )\
\int_{-\infty}^{\infty}\ \frac{ \rd x}
{F_\nu(x+\frac{\ri\pi}{2})\,F_{\nu}(-x-\frac{\ri\pi}{2})}\ ,
\eea
where $\Sigma$ stands for the regularized sum of $(-2z_j)$:
\bea\label{iasissaasuy}
\Sigma=
2\ \sum_{n=0}^{\infty}\ \Big( \,
{ \frac{ 
 \nu }{ n+1}}-z_n- z_{-n-1}\,\Big)-2\nu\ \log
\Big(\frac{r\re^{\gamma_E}}{ 4\pi}\Big)
\eea
($\gamma_E$ is   Euler's constant).
Note that 
$\Sigma=-2\ri\, \lim_{r\to 0}\zeta'(0)$, and hence    Eqs.\,\eqref{spssyssta},\,\eqref{osaosasao} imply  that
\bea\label{siaassyua}
\Sigma= \int_r^{\infty}\frac{\rd t}{t}\  \Big(\frac{\partial a}
{\partial \nu}\Big)_t\ .
\eea

\subsection{Connection problem for  the  Painlev${\acute {\bf e}}$ III equation}

We now  turn to  the ShG equation\ \eqref{luuausay} at the minisuperspace  limit.
In this   limiting  situation  the triangle $AB{\tilde B}$ in  Fig.\ref{fig1a}b  shrinks to  a segment while
 ${\hat\eta}$  becomes  a  certain solution of the Painlev${\acute {\rm e}}$ III equation\ \eqref{soisiasai}
\bea\label{mznzhx}
\lim_{\alpha\to\infty\atop
r,\,l-{\rm  fixed}}{\hat\eta}(w,{\bar w})=U(4\, |w-w_A|)\ \ \ \ \ \ \ \ \ \ \ \  
\big(\, 0<|w-w_A|< r/4\,\big)\ ,
\eea
such that
\bea\label{sksisauas}
\re^{2U(t)}=\kappa^2\ t^{4\nu-2}+o(t^{4\nu-2})\ \ \ \ \ \ \  {\rm for}\ \ \ \ \  t\to 0\ .
\eea
Parameters $\nu$ and $\kappa$ are related to $l$ and $\eta_A$\ \eqref{eaisusy} as follows
\bea\label{iiuuwwq}
\nu&=&l+{\textstyle \frac{1}{2}}\\
\log\kappa&=& -l\,  \log(4)+\lim_{\alpha\to \infty\atop
l,\, r-{\rm fixed}}\eta_A\ .\nonumber
\eea

Let us discuss  the solution satisfying \eqref{sksisauas} in the context of the
general theory of  Painlev${\acute {\rm e}}$ III equation.
For   $0<\nu<1$ and real  $\kappa$,
the  asymptotic condition\ \eqref{sksisauas} unambiguously specifies  a   two-parameter  family
of real  solutions  of the   Painlev${\acute {\rm e}}$ III equation.
A systematic  small-$t$  expansion of \eqref{sksisauas} has the form of the  double  series
\cite{Zamolodchikov:1994uw}
\bea\label{utsosaisa}
\re^{2U(t)}=\kappa^2\ t^{4\nu-2}+
 \sum_{m,n=0\atop m+n>1}^{\infty}\
B_{m,n}\ t^{ 4(1-\nu) m +4\nu n-2}
\eea
whose coefficients are uniquely determined
through  the parameters $\nu$ and
$\kappa$  by a   recursion relation   which follows from the
differential equation \eqref{soisiasai}.

The expansion\ \eqref{utsosaisa} is expected to converge for sufficiently  small $t$.
Let $t=r$  be the
closest  singularity to the origin.
The differential equation  \eqref{soisiasai} possesses the Painlev${\acute {\rm e}}$ property which states
that, except at  $t=0$ and $t=\infty$, the only possible  singularities  of $\re^{2U}$
are the  second order poles of the form
$\re^{2U(t)}=\frac{4}{(t-r)^2}-
\frac{4}{r\,(t-r)}+C+o(1)$
with some constant $C$.
Further terms in this 
Laurent expansion   are  expressed in terms of  $r$ and $C$.  They     can be easily generated
through the differential equation\ \eqref{soisiasai}.
It will  be convenient for us to substitute  $C$
for  an equivalent parameter $c$ such that
\bea\label{skkss}
\re^{2U(t)}=\frac{4}{(t-r)^2}-
\frac{4}{r\,(t-r)}+\frac{13-16\, c}{3\, r^2}+ \frac{2\, ( 16\, c-7)}{3\, r^3}\ (t-r)+O\big((t-r)^2\big)\ .
\eea
We will focus on the case when   the closest pole  to the origin   is located at  the positive real axis, i.e., $r>0$.
This requirement  imposes    certain constraints
on admissible values of  $\kappa$ and $c$.
Within the admissible  domains,
each  pair   $(\nu,\kappa)$ or  $(r, c)$ can serve  as  an independent  set of parameters
for the two-parameter  family  of  real solutions of  the Painlev${\acute {\rm e}}$ III equation
which are  regular at the segment  $t\in (0,r)$ and characterized by the behaviour $U\to (2\nu-1)\,\log(t)+O(1)$ as $t\to 0$;
However,
it is more convenient to  choose   $(r,\nu)$ with $r>0$, $0<\nu<1$, as  a basic set  of independent parameters.

At this point, we turn to the problem of  finding the functions
$\kappa=\kappa(r,\nu)$ and $c=c(r,\nu)$, i.e.,  the connection  problem 
for the local expansions
\eqref{utsosaisa} and \eqref{skkss}.
It is relatively easy to establish
the relation
\bea\label{naahsg}
 \Big(\frac{\partial c}{\partial \nu}\Big)_r=- r\,\Big(\frac{\partial \log\kappa}{\partial r }\Big)_\nu\ .
\eea
The  proof  is similar to our  previous  derivation of the
generalized FSZ relations;
One  should consider
the action functional
\bea\label{aosioas}
{\cal S}[U]=\frac{1}{4} \lim_{\epsilon\to 0}\bigg[
 \int_\epsilon^{r}\rd t\,
t \Big(   {\dot U}^2+\sinh^2(U)-\frac{2}{(r-t)^2}
 \Big)+2 (2\nu-1)\, U(\epsilon)
-(2\nu-1)^2 \log(\epsilon) \bigg]
\eea
where  the dot stands for the $t$-derivative.
For  $U(t)$,\ $t\in(0, r)$ satisfying the boundary conditions \eqref{sksisauas},\,\eqref{skkss},
the functional ${\cal S}[U]$ is well defined  and its variation vanishes
provided   $U(t)$  satisfies   the Painlev${\acute {\rm e}}$ III equation and $\delta U(r)=0$.
Let ${\cal S}^*$ be the on-shell value of \eqref{aosioas}.
One can show that
\bea\label{sosaiuas}
r\, \Big(\frac{\partial {\cal S}^*}{\partial r}\Big)_\nu&=&\frac{1}{4 }-\frac{r^2}{8}-c \\
 \Big(\frac{\partial {\cal S}^*}{\partial \nu}\Big)_r&=&\log\kappa\ ,\nonumber
\eea
and the  compatibility of these equations implies\ \eqref{naahsg}.

\begin{figure}
\centering
$(a)$\ \ 
\includegraphics[width=7  cm]{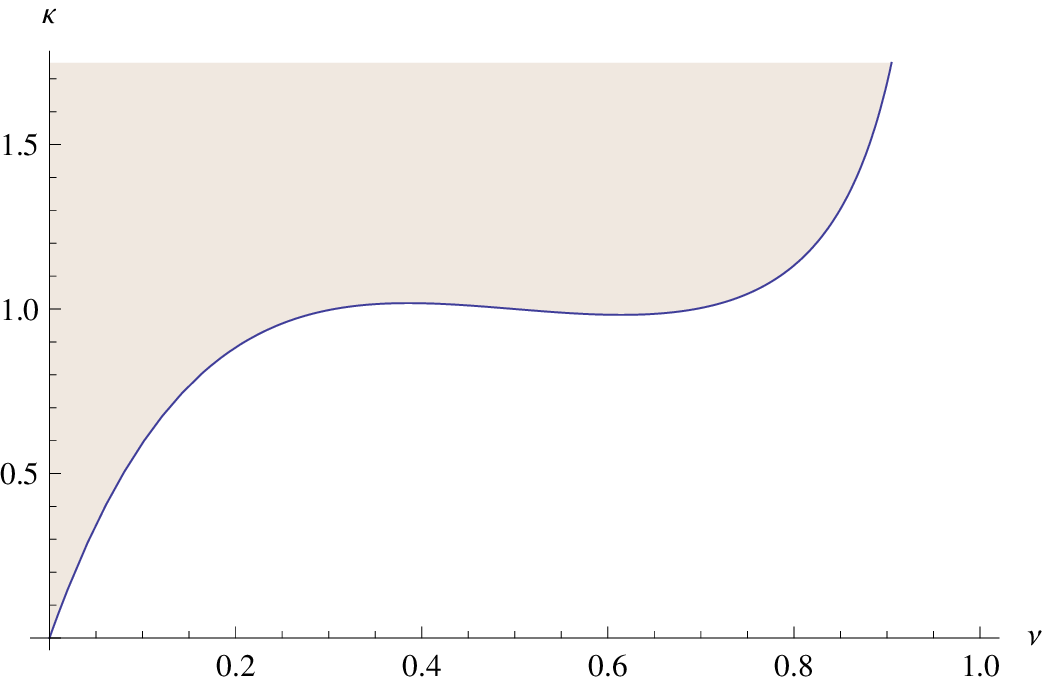}
\hskip 1.cm (b)\ \ 
\includegraphics[width=7  cm]{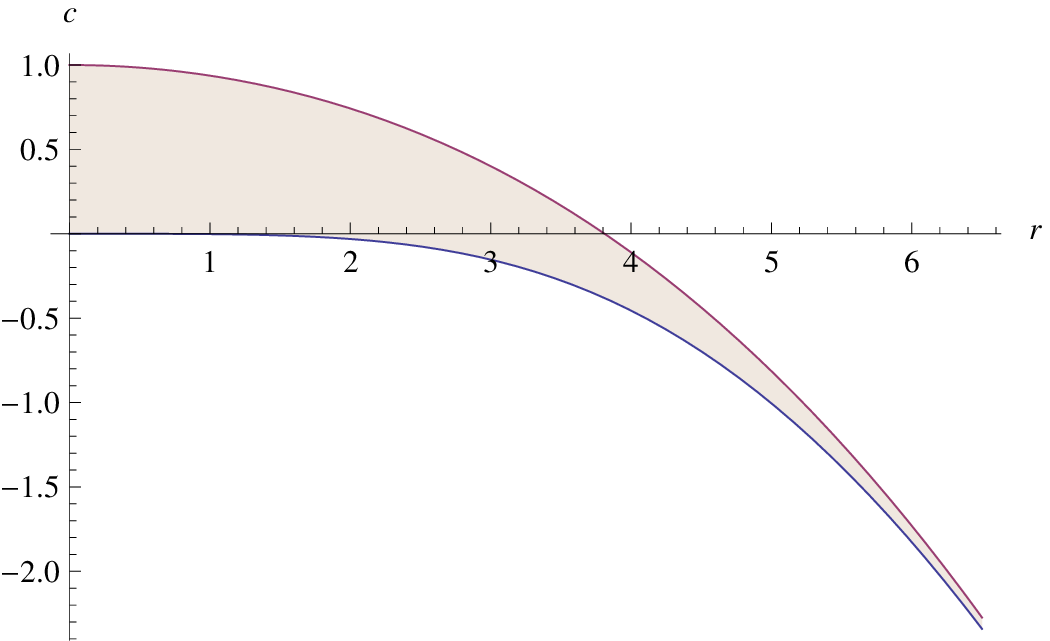}
\caption{  The admissible  parameter domain 
for the two-parameter  family of
solutions  of the   Painlev${\acute {\rm e}}$ III equation:
$(a)$  in terms of    $(\nu,\kappa)$ from  \eqref{sksisauas}. $(b)$  in terms of    $(r,c )$
from\ \eqref{skkss}. }
\label{fig5g}
\end{figure}

Now let  us  apply the results of  the previous sections.
At the  minisuperspace limit the
first generalized  FSZ relation\ \eqref{ososaasi} yields  the formula identical to\ \eqref{naahsg} with $c(r, \nu)$  
replaced by $a(r,\nu)$.
Therefore, we conclude that $c-a$ does not depend on $\nu$.
It is 
straightforward to analyze $r\to 0$ behaviour of the on-shell action:
\bea\label{jayast} {\cal S}^*= \Big(\,\frac{1}{4}-\nu^2\,\Big)\ \log(r)+\nu\log\Big(\frac{8\nu}{\re}\Big)
-\frac{1}{4}\ \log\Big(\frac{4}{\re}\Big) -\frac{r^2}{16}+O(r^4)\ .
\eea
This asymptotic formula,
combined with  \eqref{sosaiuas}, implies that $c=\nu^2+O(r^4)$,  and hence
$c-a=O(r^4)$ (see Eq.\eqref{sisaasu}). 
In  Appendix\,\ref{AppendixD}  it is explained how to systematically  recover  the   small-$r$ expansion of $c(r,\nu)$.
The calculations  yield the  expansion
\bea\label{skssia}
c&=&\nu^2+\frac{1}{2\,( \nu^2-1)}\ \Big(\frac{r}{4}\Big)^{4}+
\frac{ 5\, \nu^2+7}{ 32\, (\nu^2-1)^3\, (\nu^2-4)}\ \Big(\frac{r}{4}\Big)^{8}\\
&+&
\frac{9\, \nu^4+58\, \nu^2+29}{ 64\,  (\nu^2-1)^5\,  (\nu^2-4)\, (\nu^2-9)}\ 
 \Big(\frac{r}{4}\Big)^{12}+O(r^{16})\ ,\nonumber
\eea
which matches  exactly  the  small-$r$ expansion of $a(r,\nu)$
(see Eq.\eqref{sisaasu} and  formula 20.3.15 in Ref.\cite{Stegun}).  
The formal derivation of the relation
\bea\label{nsassysa}
c=a(r,\,\nu)
\eea
can be obtained with the use of  equations \eqref{ossai}, \eqref{uyytsoisai}
 from Appendix\,\ref{AppendixA}, 
combined with the results from Section\,\ref{MiniQ}.\footnote{
Note that      Eq.\eqref{nsassysa} allows one to derive the following  large-$\alpha$ asymptotic formula
for  the  on-shell action \eqref{ssiisa}:
$${\cal A}^*=\frac{1}{6}\ \log\Big(\, 
\frac{ 3^{\frac{1}{2}}\,  A_G^{18}\,  r}{2^4\, \alpha}\, \Big)+\frac{1}{\alpha}\ 
\bigg( {\cal S}^*+\frac{1}{4}\ \log\Big(\,\frac{ 2^{8l^2-2}\, A_G^{12}}{ \re^{\frac{4}{3}}\,  \alpha\, r}\,\Big)\, \bigg)
+o\big(\alpha^{-1}\big)\, .$$}
Eq.\eqref{nsassysa}, together with
\bea\label{saoisa}
\kappa
&=& 8\nu\ r^{-2\nu}\ \exp\bigg[\,
\int_0^{r}\frac{\rd t}{t}\ \bigg(\,2\nu- \Big(\frac{\partial a}
{\partial \nu}\Big)_t\,\bigg) \, \bigg]\\
&=&8^{1-2\nu}\ \   \frac{\Gamma(1-\nu)}{\Gamma(\nu)}\
 \exp\bigg[\,\int_r^{\infty}\frac{\rd t}{t}\  \Big(\frac{\partial a}
{\partial \nu}\Big)_t\,\bigg]    \ ,\nonumber
\eea
leads to an  explicit solution  of  the connection problem for   local expansions\ \eqref{utsosaisa} and
\eqref{skkss}.
The domain of  applicability of  Eqs.\eqref{nsassysa},\,\eqref{saoisa}
is given by the inequalities
(see Fig.\,\ref{fig5g}a) 
\bea\label{asosiisaa}
0<\nu< 1\ ,\ \ \  \ \ \ \ \kappa>8^{1-2\nu}\ \   \frac{\Gamma(1-\nu)}{\Gamma(\nu)}\ .
\eea
It can be equivalently described
in terms of the pair   $(r,c)$: 
\bea\label{akjasjwqi}
r>0\ ,\ \ \ \ \ \ \ \ \ \ a_0(r)< c< b_1(r)\ ,
\eea
where $a_0=a(r,0)$ and  $b_1=\lim_{\nu\to 1}a(r,\nu)$ stand for  the minimum and maximum of
the first conduction band for the Mathieu equation, respectively (see Fig.\,\ref{fig5g}b).

Finally, let us briefly discuss the minisuperspace limit for  $\Phi^{(-)}=\frac{1}{2}\, (\,\Phi-{\hat \eta})$.
Contrary to ${\hat \eta}(w,{\bar w})$, the potential $\Phi(w,{\bar w})$  does not have  a  finite limit for   $0<|w-w_A|< r/4$.
However,   the   divergent part of    $\Phi^{(-)}(w,{\bar w})$  for   $0<|w-w_A|< r/4$
is somewhat  trivial and can    be  resolved by by means of the decomposition
\eqref{rwsosaosai} from Appendix\,\ref{AppendixA}.
The non-trivial part  ${\tilde \Phi}^{(-)}$
is specified by the  conditions\ \eqref{sosiosai},\eqref{usysosiosa} 
and remains finite  as $\alpha\to \infty$.
It is convenient   to introduce
\bea\label{sosisaiu}
W(t)=
-l (l+1)\ \log|w-w_A|+
\lim_{\alpha\to\infty\atop
r,\,l-{\rm  fixed}} {\tilde \Phi}^{(-)}(w,{\bar w})\ ,
\eea
where $t=4\,|w-w_A|<r$ and $l=\nu-\frac{1}{2}$.
As follows from  Eqs.\eqref{sksai},\, \eqref{rwsosaosai}, $W(t)$
satisfies  the  linear inhomogeneous differential equation (assuming  $U$ is given)
\bea\label{siksisasai}
\frac{4}{t}\ \frac{\rd }{\rd t}\Big(\, t\, \frac{\rd W }{\rd t}\,\Big)= \re^{-2U(t)}-1\ ,
\eea
and the boundary condition 
\bea\label{xskisai}
W(t)=
\Big(\,\frac{1}{4}-\nu^2\,\Big)\ \log\Big(\frac{t}{4}\Big) +o(1)\ ,\ \ \ {\rm as}\ \ \  \ \ t\to 0\ .
\eea
Therefore
\bea\label{ysosaisa}
W(t)=\Big(\,\frac{1}{4}-\nu^2\,\Big)\ \log\Big(\frac{t}{4}\Big)
-\frac{t^2}{16}-\frac{1}{4}\ \int_0^t\rd \tau\,  \tau\, \log\Big(\frac{\tau}{t}\Big)\  \re^{-2U(\tau)}\ .
\eea
The  $t$-derivative
of ${ W}(t)$ at  $t=r$ is given by
\bea\label{xisiiauysr}
r\,{\dot W}(r)=\frac{1}{4}-\frac{r^2}{8}-\nu^2+\frac{1}{4 }\ \int_0^r\rd t\,  t\,  \re^{-2U(t)}
\eea
and it is not difficult to show  that
\bea\label{lsissu}
{\dot W}(r)=\Big(\frac{\partial {\cal S}^*}{\partial r}\Big)_\nu
\ .
\eea
The last two  relations
combined with   Eqs.\eqref{sosaiuas},\,\eqref{nsassysa}   imply\label{isiiauysr}
\bea\label{aiassayas}
\frac{1}{4 }\ \int_0^r\rd t\,  t\,  \re^{-2U(t)}=\nu^2-a(r,\nu)\ .
\eea
By transforming the integrals as in Ref.\cite{Wu:1975mw},
the   value of  $W(t)$ at $t=r$ can be  expressed in terms of $a$, $\kappa$ and  ${\cal S}^*$:
\bea\label{isuusasy}
W(r)=\frac{1}{2}\ \log\Big(\frac{r}{4}\Big)-\frac{r^2}{8}+
\Big(\nu-\frac{1}{2}\Big)^2
+2\nu^2\,\log(2)-a
+\nu\ \log(\kappa)-{\cal S}^*\ ,
\eea 
which is equivalent to the following relations
\bea\label{usossai}
W(r)=\Big(\, \frac{1}{4}-\nu^2\,\Big)\, \log\Big(\frac{r}{4}\Big)- 
\frac{r^2}{16}-
\int_0^{r}\frac{\rd t}{t}\, \bigg(\,
t\, \Big(\frac{\partial a}
{\partial t}\Big)_\nu+
\nu\, \Big(\frac{\partial a}
{\partial \nu}\Big)_t-a(t,\nu)-\nu^2\, \bigg)\ , 
\eea
or
\bea\label{sossisosisai}
W(r)&=&{\cal Y}_0(\nu)-
\log\bigg[\,   A_G^3\ \ \ \re^{-\frac{1}{4}-\nu^2}\   2^{2\nu^2+\frac{2}{3}}\ 
\bigg(\,  \frac{\Gamma(1+\nu)}{\Gamma(1-\nu)}\,\bigg)^{\nu}\,\bigg]\\
&+&
\int_r^{\infty}\frac{\rd t}{t}\ \bigg(\,
t\, \Big(\frac{\partial a}
{\partial t}\Big)_\nu+
\nu\, \Big(\frac{\partial a}
{\partial \nu}\Big)_t-a(t,\nu)
+\frac{t^2}{8}-\frac{1}{4} \,\bigg)\ ,\nonumber
\eea
where ${\cal Y}_0(\nu)$ is given by Eq.\eqref{tskssia}.
Note that the last term in Eq.\eqref{sossisosisai}  vanishes as $r\to\infty$ while  the remaining part 
reproduces the result  quoted in  Ref.\cite{Zamolodchikov:1994uw}.


\section{Concluding remark}

In this work we have described the link  between the action functional for the classical
ShG equation and the  YY-function corresponding to the $k$-vacuum states  in the quantum sine-Gordon model.
The natural question arises: Can this relation be generalized for the excited states?
Nowadays the machinery of the Destri de Vega equation for the excited states are well developed 
\cite{Destri:1997yz, Feverati:1998dt}, so that  the  calculation of  the  YY-function for
the excited states does not seem to be a  particularly complicated problem.
However, it remains unclear how to construct  integrable classical equations
associated with the  excited states and perhaps more importantly,  what all of this really means.

\section*{Acknowledgments}

Numerous discussions with  A.B. Zamolodchikov were  highly valuable for me.
I also want to acknowledge helpful   discussions  with V. Bazhanov,  N. Nekrasov, S. Shatashvili and F. Smirnov.
\bigskip

\noindent This  research was supported in part by DOE grant
$\#$DE-FG02-96 ER 40949.

\appendix

\renewcommand{\theequation}{\thesection.\arabic{equation}}

\section{\label{AppendixA} Appendix:  Basic properties of the  potential  $\Phi$ }

Let us define  the following  linear  combinations   of
the solution   ${\hat\eta}$  of the ShG equation  \eqref{luuausay}  and  the potential $\Phi$  
\eqref{ystsossai} 
\bea\label{sossaoi}
\Phi^{(\pm )}={\textstyle\frac{1}{2}}\ \big(\,\Phi\pm{\hat \eta}\,\big)\ .
\eea
They  satisfy
the relations
\bea\label{sksai}
\partial^2_{w}\Phi^{(\pm )}=
{\textstyle \frac{1}{2}}\ \big(\, (\partial_w{\hat\eta})^2\pm \partial^2_w{\hat\eta}\,\big)\, ,\ \ \
\partial^2_{\bar w}  \Phi^{(\pm )}=
{\textstyle \frac{1}{2}}\ \big(\,(\partial_{\bar w}{\hat\eta})^2\pm \partial^2_{\bar w}{\hat\eta}\,\big)\ ,\ \ \
\
\partial_{w}\partial_{\bar w}\Phi^{(+)}=\re^{\pm 2{\hat \eta}}-1\, ,
\eea
which can be considered as a closed system of nonlinear  partial differential equations.
Results of Refs.\cite{Zamolodchikov:1994uw} and \cite{Lukyanov:2010rn} imply
that the desirable  solution of  \eqref{sksai} 
is  expressed in terms of  
 the  Fredholm determinants
\bea\label{ossai}
\Phi^{(\pm)}(w,{\bar w})=\log \det\bigg(\,{\mathbb I}\pm \frac{{\mathbb K}_{w,{\bar w}}}{4\pi}\, \bigg)\ .
\eea
The  kernel  of the integral  operator ${\mathbb K}_{w,{\bar w}}$ reads
\bea\label{soisia}
K_{w,{\bar w}}(\theta,\theta')&=&
\frac{  \sqrt{g_{w,{\bar w}}(\theta) g_{w,{\bar w}}(\theta')}}{ \cosh\big(\frac{\theta-\theta'}{ 2}\big)}\ ,\\
g_{w,{\bar w}}(\theta)&=&  T_{\frac{1}{2}}
\big(\theta+{\textstyle \frac{\ri\pi (\alpha+1)}{2\alpha}}\big)\ 
\exp\big( -2 w\, \re^{\theta}  -2{\bar  w}\, \re^{-\theta}\, \big)\ ,\nonumber
\eea
where  $ T_{\frac{1}{2}}(\theta)$ is defined by
the  $Q$-function\ \eqref{saossisa} through  Baxter's 
equation \eqref{isssisa} with $\xi=\alpha^{-1}$.
In \eqref{ossai}  $(w,{\bar w})$ play a r${\hat {\rm o}}$le of complex  parameters. 
Within   the domain 
\bea\label{sosisa}
|\arg(w)|\leq {\textstyle \frac{\pi (\alpha+1)}{2\alpha}}\ ,
\eea
$\Phi^{(\pm)}(w,{\bar w})$ can be represented by convergent series
\bea\label{saisa}
&&\Phi^{(\pm)}(w,{\bar w})=\sum_{n=1}^\infty\frac{(\pm 1)^n}{n}\ \Phi_n(w,{\bar w})\ ,\\
&& \Phi_n(w,{\bar w})=\int_{-\infty}^\infty\prod_{j=1}^{n}
\left(\, \frac{\rd\theta_j}{4\pi}\
\frac{ g_{w,{\bar w}}(\theta_j)}
{\cosh\big(\frac{\theta_j-\theta_{j+1}}{ 2}\big)}\, \right)\ ,\nonumber
\eea
where it is implied that $\theta_{n+1}\equiv\theta_1$.
Some immediate consequences of \eqref{saisa},
\bea\label{xssisaias}
\Phi^{(\pm)}\big( w\,\re^{ \frac{\ri \pi (\alpha+1)}{\alpha}},\,  
{\bar w}\,\re^{-\frac {\ri \pi (\alpha+1)}{\alpha}}\,\big)&=&
\Phi^{(\pm)}(w,\,{\bar w)}
 \ \ \ \ \ \ \ \ {\rm for}\ \ \ \ \ |w|>|w_B|
\eea
and  
\bea\label{saisisua}
\lim_{|w|\to\infty}\Phi^{(\pm)}(w,{\bar w})=0\ ,
\eea
have been used in the derivation of basic Eq.\eqref{xososa}.
By deforming the integration contours in \eqref{saisa}, the 
applicable domain \eqref{sosisa}  can be slightly extended to the triangle $OB{\tilde B}$  (see  Fig.\ref{fig1a}b). 
This leads to the fact that for $ \Re e( w-w_B)\geq 0$
\bea\label{ksaisai}
\Phi^{(\pm)}(w,{\bar w})= \Phi^{(\pm)}({\bar w}, w )\ .
\eea
Yet   the expansions  \eqref{saisa} cannot be applied inside the triangle $AB{\tilde B}$,
\bea\label{soosa}
0<\Re e (w-w_{A})< \Re e (w_{B}-w_{A})\,,\ \ \ \ \ \  \ \arg(w-w_{A})<\frac{\pi}{2\alpha}\ .
\eea
Within this  domain,  ${\hat \eta}(w,{\bar w})$ satisfies the symmetry  relations
\bea\label{xsosaosao}
{\hat \eta}\big( w_A+   (w-w_A)\, \re^{\frac{\ri \pi }{\alpha}},\, {\bar w}_A+
({\bar w}- {\bar w}_A)\, \re^{-\frac {\ri \pi }{\alpha}}\,\big)
=
{\hat \eta} (w,\,{\bar w})={\hat \eta} ({\bar w}, \, w)\ ,
\eea
whereas,  as follows from \eqref{sksai},
\bea\label{xsosisai}
\partial_w^2\, \delta \Phi^{(\pm)}=
\partial_{\bar w}^2\,\delta \Phi^{(\pm)}=\partial_w\partial_{\bar w}\,\delta \Phi^{(\pm)}=0\ ,
\eea
where
$\delta \Phi^{(\pm)}= \Phi^{(\pm)}\big( w_A+   (w-w_A)\, \re^{\frac{\ri \pi }{\alpha}},\, {\bar w}_A+
({\bar w}- {\bar w}_A)\, \re^{-\frac {\ri \pi }{\alpha}}\,\big)
-
\Phi^{(\pm)}(w,\,{\bar w})$. 
From  Eqs.\eqref{skssai},\,\eqref{xsosisai}
and the reflection symmetry relation \eqref{ksaisai}  
applied to the segment $B{\tilde B}$, we  deduce  that within the domain \eqref{soosa} 
$\Phi^{\pm}$ can be written  in  the form
\bea\label{rwsosaosai}
\Phi^{(\pm)}(w,{\bar w})=-l(l \mp  1) \, \log|w-w_{A}|+ (w-w_A)\, {\bar J}+
({\bar w}- {\bar w}_A)\, J+
\Phi^{(\pm)}_A+ {\tilde  \Phi}^{\pm}(w,{\bar w})\, ,
\eea
where $\Phi^{(\pm)}_A$ and  $J={\bar J}$ are  some real constants, 
\bea\label{sosiosai}
{\tilde {\Phi}}^{(\pm)}\big( w_A+ (w-w_A)\,\re^{ \frac{\ri \pi }{\alpha}},\, 
 {\bar w}_A+ ({\bar w}- {\bar w}_A)\,\re^{-\frac {\ri \pi }{\alpha}}\,\big)
={\tilde  \Phi}^{\pm} (w,\,{\bar w})={\tilde  \Phi}^{\pm} ({\bar w},\, w)
\eea
and
\bea\label{usysosiosa}
\lim_{|w-w_A|\to 0} {\tilde {\Phi}}^{(\pm)}(w,{\bar w})=0\ .
\eea

The constants $\Phi^{(\pm)}_A$  in \eqref{rwsosaosai} are given by the
linear combinations of ${\hat \eta}_A$\ \eqref{eaisusy} and  $\Phi_A$\ \eqref{isusy}:
\bea\label{saossia}
\Phi_A^{(\pm)}={\textstyle \frac{1}{2}}\, (\, \Phi_A\pm \eta_A\,)\ .
\eea
As for  $J$ and ${\bar J}$, 
the calculations performed in Section\,\ref{Section2b} 
imply that\footnote{
For $r=0$,\ equation \eqref{uyytsoisai} (supplemented by \eqref{skasisaus},\,\eqref{sksjsasayusa},\,\eqref{uyast})  
follows immediately from  the results of Ref.\cite{Zamolodchikov:1994uw}. It 
was  used
in Ref.\,\cite{Alday:2009yn} for  the case $l=0$ and  $2\alpha=2,\,3\ldots $\ .}
\bea\label{uyytsoisai}
\int_{D}{ \frac{{\rd w}\wedge {\rd }{\bar w}}{2\pi \ri}}\ \ 4\,\big(\, \re^{\pm 2 {\hat \eta}}-1\,\big)=
\frac{l(l\mp 1)  }{\alpha}\pm\frac{1 }{2}+ \frac{1}{12}-\frac{r}{2\pi}\ 
\sin\Big({ \frac{\pi}{2\alpha}}\Big)\  \big(\,  {\mathfrak J}_1+ {\bar   {\mathfrak J}}_1\,\big)
  \ .
\eea
On the  other hand, the integrand here
coincides with  the Laplacian  $4\,\partial_w\partial_{\bar w}\Phi^{(\pm)}$, and
the L.H.S. of \eqref{uyytsoisai} can be  expressed in terms of  $J={\bar J}$.
This yields the relations
\bea\label{uosisai}
J= {\mathfrak J}_1\ ,\ \ \ \ \ \ \ \ \ \ \ \ \ {\bar J}= {\bar   {\mathfrak J}}_1\ .
\eea

Finally, note  that   Eqs.\eqref{skssaisusy},\,\eqref{sksai} yield
the following asymptotic behaviours   in the vicinity of  $B\sim {\tilde B}$  (see  Fig.\,\ref{fig1a}b)
\bea\label{skssusysai}
{ \Phi}^{(+)}\to -\frac{7}{36}\ \log|w-w_{B}|+O(1)\ ,\ \ \  \  \ \ \ \ \
{ \Phi}^{(-)}\to \frac{5}{36}\ \log|w-w_{{\tilde B}}|+O(1)\ .
\eea

\section{\label{AppendixB} Appendix:  Calculation of ${\cal A}^*_{\infty}=\lim_{r\to\infty}{\cal A}^*$}

For large $r$   the dominating contributions to  the on-shell action\ \eqref{ssiisa} come from the
vicinities  of the points  $A$ and $B\sim {\tilde B}$ (see Fig.\ref{fig1a}b). 
Near these points,  ${\hat\eta}$ can be approximated by
$U(4|w-w_A|)|_{\nu=l+\frac{1}{2}}$ and $U(4|w-w_{B}|)|_{\nu=\frac{1}{3}}$, respectively,  where
$U$ are the   Painlev${\acute {\rm e}}$ III  transcendents, i.e., 
regular at $t>0$   solutions of Eq.\eqref{soisiasai} satisfying the boundary conditions
\bea\label{saksaisaiu}
U\to
\begin{cases}
&(2\nu-1)\ \log\big(\frac{t}{4}\big)+U_0(\nu)+o(1)\ ,\ \ \ \ {\rm as}\ \ \ \ t\to 0\nonumber\\
& 0\ ,\ \  \ \ \ \ \ \ \ \ \ \ \   \ \ \ \ \ \ \ \ \ \ \ \ \ \ \ \ \ \ \  \ \   \qquad  \ \ \ \  {\rm as}\ \ \ \ t\to \infty
\end{cases}
\ .
\eea
This observation implies that the limiting value of the on-shell action \eqref{ssiisa}
is given by
\bea\label{osoasos}
{\cal A}^*_\infty=
3\ 
{\cal S}^*_\infty\big({\textstyle {\frac{1}{3}}}\big)+
\alpha^{-1}\ {\cal S}^*_\infty(\nu)\ ,
\eea
where ${\cal S}^*_\infty(\nu)$ is the on-shell value of the
action functional  for the Painlev${\acute {\rm e}}$ III equation,
\bea\label{ssisa}
{\cal S}_\infty[U]=\frac{1}{4}\ \lim_{\epsilon\to 0}\ \bigg[\,
\int_{\epsilon}^\infty\rd t\ t\ \big(\, {\dot U}^2+\sinh^2(U)\, \big)+
2\, (2\nu-1)\ U(\epsilon)- (2\nu-1)^2\ \log\Big(\frac{\epsilon}{4}\Big)\, \bigg]\, .
\eea
Here  the dot stands for  the $t$-derivative.
Varying  \eqref{ssisa}   with respect to
$\nu$ yields equation
$U_0(\nu)= \frac{{\rm d} {\cal S}^*_\infty}{{\rm d}\nu}$.
Since    ${\cal S}^*_\infty(\frac{1}{2})=0$, one has
\bea\label{sisisia}
{\cal S}^*_\infty(\nu)=\int_{\frac{1}{2}}^{\nu}\rd x\ U_0(x)\ .
\eea
An explicit form of  $U_0(\nu)$  is well known\ \cite{McCoy:1976cd},
\bea\label{salaoiasio}
U_0=\log\left(\,2^{1-2\nu}\, \frac{\Gamma(1-\nu)}{
\Gamma(\nu)}\, \right)\ .
\eea
Combining  \eqref{osoasos},\,\eqref{sisisia} with \eqref{salaoiasio} one arrives to \eqref{ssisaisa}.

\section{\label{AppendixC} Appendix:   Formula  for $\big(\frac{\partial{\mathfrak Y}}{\partial \xi}\big)_{r, k}$}

This section presents some  supporting evidences for relation \eqref{assjsajasju} which is expected to hold for any $\xi>0$.

Using Eqs.\eqref{soisoai}, \eqref{sksklsskla} and \eqref{usystsoissisa}
it is easy to show that
for $\xi\geq 1$  (i.e., in the case without  soliton-antisoliton bound states)
the  large-$r$ behaviour  of ${\mathfrak Y}$
is given  by
\bea\label{saossai}
{\mathfrak Y}(r)=-\frac{2}{\pi^2}\ \cos(2\pi k)\ K_0(r)+O( \re^{-2r})\ ,
\eea
whereas
\bea\label{sosaasisau}
\zeta(\omega)=-
\frac{\omega\, \sinh(\frac{\pi\omega (1+\xi)}{ 2})}{\pi\,
 \cosh(\frac{\pi\omega}{ 2})\,
\sinh(\frac{\pi\xi \omega}{ 2})}\ \cosh\Big(\,\frac{\pi\omega}{2}-2\pi\ri k\Big)\  K_{\ri\omega}(r)+O(\re^{-2r})\ .
\eea
Here $K_\nu(r)$ is the modified Bessel function and  the symbol $O(\re^{-2r})$ 
stands for  an asymptotic behaviour of the form $\propto r^\gamma\,\re^{-2r}$ as $r\to\infty$
with  some exponent $\gamma$.
The leading large-$r$  asymptotic of ${\mathfrak Y}$ does not depend on $\xi$ and hence
the  L.H.S. of \eqref{assjsajasju} decays as  $O(\re^{-2r})$.  Thus Eq.\eqref{assjsajasju} 
is qualitatively consistent 
with asymptotics\ \eqref{saossai} and \eqref{sosaasisau}.
A quantitative comparison can be made
for $\xi=2$ and $k=\pm \frac{1}{4}$.
In this case\ \cite{Zamolodchikov:1994uw} 
\bea\label{ssasaaussau}
\Big(\frac{\partial{\mathfrak Y}}{\partial \xi}\Big)_{r, k}=
\frac{1}{2\pi^2}\ \int_{r}^\infty\rd t\ t\, \log\Big(\frac{t}{r}\Big)\, K_0^2(t)+O(\re^{-4r})\ 
\eea
(This formula follows immediately  from\ \eqref{siasai} and the large-$t$ 
asymptotic $U(t)=\frac{1}{\pi}\, K_0(t)+O(\re^{-2t})$), so that
Eq.\eqref{assjsajasju} implies
the following  easily established identity
\bea\label{suusaausyt}
 \int_{r}^\infty\rd t\ t\, \log\Big(\frac{t}{r}\Big)\, K_0^2(t)=
\int_{-\infty}^\infty\rd\omega\  \, \frac{\omega\ K^2_{\ri\omega}(r)}{2\sinh(\pi\omega)}\ .
\eea

Another piece of  evidence  supporting  \eqref{assjsajasju} comes   from the  consideration of $r\to 0$ limit.
Using Eq.\eqref{oaasauas}  one finds  
\bea\label{nasgsat}
\frac{1}{4}\ \int_{-\infty}^\infty
\frac{\rd\omega}{\omega}\
\frac{ \sinh(\pi \omega)}
{ \sinh^2(\frac{\pi\omega(1+\xi)}{2})}\ \zeta(\omega)\,
\zeta( -\omega)+\frac{2\ri\,  k}{(1+\xi)^2}\ \zeta'(0)=F(k,\xi)+F(-k,\xi)+o(1)\ ,
\eea
where $F(k,\xi)$ stands for
\bea\label{siisa}
F(k,\xi)=\frac{1}{4}\ \  \dashint_{-\infty}^\infty
\frac{\rd\omega}{\omega}\
\frac{ \sinh(\pi \omega)}
{ \sinh^2(\frac{\pi\omega(1+\xi)}{2})}\  \zeta^{\rm(cft)}_k(\omega)\,\zeta^{\rm(cft)}_k(-\omega)\ .
\eea
As follows from results of Refs.\cite{Bazhanov:1996dr,Bazhanov:1998wj},
$\zeta^{\rm(cft)}_k(\omega)$ 
is a  meromorphic   function of the complex variable $k$,
analytic in the half plane $\Re e(k)>  -\frac{1}{2}$. 
Also, the product
 $\zeta^{\rm(cft)}_k(\omega)\,\zeta^{\rm(cft)}_k(-\omega)$
can be represented in the form
\bea\label{isausasau}
 \zeta^{\rm(cft)}_k(\omega)\,\zeta^{\rm(cft)}_k(-\omega)= \frac{  \sinh(\frac{\pi\omega(1+\xi)}{2})}
{ 24\, \sinh(\frac{\pi\omega\xi}{2}) \cosh(\frac{\pi\omega}{2})}\ \left(\, -\frac{c_{\rm eff}}{\omega^2+1}+1
+\omega^2\, X
\,\right)\ ,
\eea
where 
$X$ admits  the    $k\to\infty$ asymptotic expansion  as  $\Re e(k)>  -\frac{1}{2}$:
\bea\label{sosisaisa}
X\asymp \sum_{n=1}^\infty X_n(\omega^2,h)\ p^{-2n }
\eea
with $p=2k \xi$ and  $h=\xi(1+\xi)$.
The coefficient $X_n(\omega^2,h) $  are calculated systematically   with the  WKB method applied to
the  Schr${\ddot {\rm o}}$dinger operator \ \eqref{xssiisa}.
They turn out  to be 
polynomials in  the variables  $\omega^2$ and $h$  
of orders $n-1$ and $n$, respectively. 
For example,
\bea\label{iasissau}
X_1(\omega^2,h)&=&{\textstyle \frac{1}{10}}\ h\\
X_2(\omega^2,h)&=&{\textstyle \frac{1}{126}}\
\Big(\, \omega^2\ h^2- 9\, h^2-{\textstyle \frac{22}{5}}\ h-{\textstyle \frac{2}{5}}\, \big)\ .
\nonumber
\eea
Thus  $F(k,\xi)$  \eqref{siisa} can be written in the form
\bea\label{qwwwe}
F( k,\xi)=-\frac{c_{\rm eff}}{24}\
\bigg[\, \psi\Big(\frac{\xi}{2}\Big)-\psi\Big(\frac{3}{2}+\frac{\xi}{2}\Big)+
\frac{1+2\xi}{\xi (1+\xi)}\, \bigg]
+\frac{1}{24}\ \log\Big(\frac{\xi}{1+\xi}\Big)+{\tilde F}\ ,
\eea
where $\psi(z)=\frac{\rd }{\rd z}\,\log\Gamma(z)$ and
${\tilde F}$ admits
 the large-$k$ asymptotic expansion
 in the half-plane $\Re e( k)>-\frac{1}{2}$
\bea\label{sosusau}
{\tilde F}\asymp\sum_{n=1}^\infty{\tilde F}_n(\xi )\ p^{-2n}\ \ \ \ \ \ \ \ \ \ (\, p=2k \xi\,)\ .
\eea
I have calculated explicitly 
the expansion coefficients  ${\tilde F}_n$ up to the order $n=8$ and found full agreement with the formula
\bea\label{qssaisiuq}
&& {\tilde F}=\Big(\,\frac{\partial}{\partial \xi}+\frac{1+2\xi }{2\xi (1+\xi)}\  \ p\, \frac{\partial}{\partial p}\,\Big)\,
\bigg[\
\frac{c_{\rm eff}}{24}\ \log\big(\,\xi(1+\xi)\,\big)-\\
&&
\int_0^\infty\frac{\rd t}{t}\
\bigg(\, \frac{\sinh( t)\ \re^{-2 p t}}{ 4t\sinh(\xi t)\sinh\big(t(1+\xi)\big)}-\frac{1-2p t}{4\xi (1+\xi) t^2}
+\frac{c_{\rm eff}}{12}\ \re^{-2  t}\, \bigg)\, \bigg]\nonumber
\eea
$\big(\,c_{\rm eff}=1-\frac{6\, p^2}{\xi(1+\xi)}\,\big)$, which follows from 
Eqs.\eqref{usystsoissisa},\,\eqref{assjsajasju},\,\eqref{nasgsat} combined with
\eqref{sisaiaissus},\,\eqref{ssisisa}.

\section{\label{AppendixD} Appendix: Small-$r$ expansion of $c(r,\nu)$ }

Here 
we  explain  how  to  develop the    expansion\ \eqref{skssia}  systematically.

The partial resummation of the double  series \eqref{utsosaisa} yields
the following structure
\bea\label{ksiis}
\re^{2U(t)}=\frac{(8\nu)^2\, X}{ t^2\,(1-X)^2}\  
\sum_{n=0}^\infty\, \frac{ P_{3n}(X\,|\,\nu) }{ X^n\,(1-X)^{n}}\  
\Big(\frac{t}{8}\Big)^{4n}\ ,
\eea
where 
$X=(\frac{\kappa}{8\nu})^2\ t^{4\nu}$,
$P_0=1$ and  $P_{3n}$ are  polynomials in $X$ with the degree ${\rm deg}(P_{3n})\leq 3n$,
such that
\bea\label{kslsakl}
P_{3n}(X\,|\,\nu)=(-1)^n\ X^{3n}\ P_{3 n}(X^{-1}\,|-\nu)\ .
\eea
With  sufficient  computer resources,  
these polynomials can be calculated  order by order with reasonable facility.
Explicitly,
\bea\label{ssausaua}
P_3(X)= \frac{2}{ \nu^2\, (1-\nu)^2}\ \ \big( \, (1-\nu)\, X-\nu-1\,\big)\, \big(\,
(1-\nu)\, X^2+(4\nu^2-2)\, X+\nu+1\,\big)
   \ .
\eea
As the next step, one should   resummate    the series\ \eqref{ksiis} and bring it to the 
form
\bea\label{skssisai}
\re^{2U(t)}=t^{-2}\  \bigg[\, \frac{V_{-2}}{(X-B^2)^2}+
\frac{V_{-1}}{X-B^2}+V_0\ t^4+\sum_{n=1}^\infty V_n\ t^{4 n+4}\ (X-B^2)^n
\, \bigg]\ ,
\eea 
where   $B=B(t)$ and
$V_j=V_j(t)\, ,\ j=-2,\,-1,\,0\ldots$  are formal  power series   in $t^4$.
Note that this form is suggested by  the Laurent series  \eqref{skkss}.
Comparing  the singular  parts     at $t=r$ of
\eqref{skssisai} ane \eqref{skkss}
one finds that
\bea\label{sisai}
\kappa=8\nu\ r^{-2\nu} \ B(r)\ ,
\eea
whereas
$V_{-2}$ and $V_{-1}$ are certain differential polynomials of $B=B(t)$:
\bea\label{skssaisa}
V_{-2}&=&
16\ B^2\ (\, 2\, \nu\, B- t\, {\dot B}\,)^2\ ,\\
V_{-1}&=&8\   \big(\, 8\, \nu^2\, B^2+ (1- 8\,\nu)\, t\, B\, {\dot B}+t^2\ {\dot B}^2+
 t^2\, B\, {\ddot B}\, \big)\ .
  \nonumber
\eea
Here the dots  stand for derivatives with respect to $t$.
To determine the   expansion coefficients of the  formal  power series $B=B(t)$ and $V_0=V_0(t)$ 
one should  re-expand  \eqref{skssisai}  in the powers  of  $(X-1)$ and   compare
the   terms $\propto   (X-1)^{-n}$  for $n\geq  0$ 
with  the  similar   terms  from   \eqref{ksiis}.
Explicit calculations yield    Eq.\eqref{skssia} and the similar expansion for $\log\kappa$. 
I verified  that  the    power series for $c$ and $\log\kappa$
obey  the relation  \eqref{naahsg}  up to  in twelfth order in $r$.

\end{document}